\documentclass[bibyear]{aa} 

\usepackage{xcolor}
\usepackage{graphicx}
\usepackage{txfonts}
\usepackage{booktabs}
\usepackage{siunitx}
\usepackage[colorlinks=true,citecolor=blue]{hyperref}
\usepackage{longtable}
\usepackage[flushleft]{threeparttable}
\usepackage{amsmath}
\usepackage{upgreek}
\usepackage{multicol}
\usepackage{multirow}
\usepackage{natbib}
\usepackage{orcidlink}
\usepackage{comment}
\usepackage{float}
\usepackage{txfonts}
\usepackage{upgreek}

\defcitealias{Giarratana2024}{G24}

\begin{document}

\title{Comparing the size evolution of GRB 221009A with afterglow models}

   \author{S. Giarratana\inst{1}\fnmsep\thanks{Corresponding author: stefano.giarratana@inaf.it}\orcidlink{0000-0002-2815-7291} \and
   O.~S. Salafia\inst{1,2}\orcidlink{0000-0003-4924-7322} \and
   B. Marcote \inst{3,4}\orcidlink{0000-0001-9814-2354} \and
   Z. Paragi \inst{3}\orcidlink{0000-0002-5195-335X} \and
   M. Giroletti\inst{5}\orcidlink{0000-0002-8657-8852} \and
   G. Ghirlanda\inst{1,2}\orcidlink{0000-0001-5876-9259}
   \and
   L. Nava\inst{1}\orcidlink{0000-0001-5960-0455}
   \and
   A. van der Horst\inst{6}\orcidlink{0000-0001-9149-6707}
    \and
   J.~K. Leung\inst{7,8,9,5}\orcidlink{0000-0002-9415-3766}
   \and
   L. Rhodes\inst{10,11}\orcidlink{0000-0003-2705-4941}
   \and
   P.~G. Edwards\inst{12}\orcidlink{0000-0002-8186-4753}
   \and
   M. P\'erez-Torres\inst{13}\orcidlink{0000-0001-5654-0266}
   \and
   J. Yang\inst{14}\orcidlink{0000-0002-2322-5232}
    }

   \institute{
   INAF Osservatorio Astronomico di Brera, Via E. Bianchi 46, I-23807 Merate, Italy
   \and 
   INFN -- sezione di Milano-Bicocca, Piazza della Scienza 3, I-20126 Milano, Italy
   \and
   Joint Institute for VLBI ERIC (JIVE), Oude Hoogeveensedijk 4, 7991PD Dwingeloo, The Netherlands 
   \and
   ASTRON, Netherlands Institute for Radio Astronomy, Oude Hoogeveensedijk 4, 7991~PD Dwingeloo, The Netherlands
   \and
   INAF Istituto di Radioastronomia, via Gobetti 101, I-40129 Bologna, Italy
   \and
   Department of Physics, George Washington University, 725 21st St NW, Washington, DC 20052, USA
   \and
   David A. Dunlap Department of Astronomy and Astrophysics, University of Toronto, 50 St. George Street, Toronto, ON M5S 3H4, Canada
   \and
   Dunlap Institute for Astronomy and Astrophysics, University of Toronto, 50 St. George Street, Toronto, ON M5S 3H4, Canada
   \and
   Racah Institute of Physics, The Hebrew University of Jerusalem, Jerusalem 91904, Israel
   \and
   Department of Physics, McGill University, 3600 Rue University, Montréal, QC H3A 2T8, Canada
   \and 
   Trottier Space Institute, McGill University, 3550 Rue University, Montréal, QC H3A 2A7, Canada
   \and
   CSIRO Space and Astronomy, Australia Telescope National Facility, PO Box 76, Epping 1710 NSW, Australia
   \and
   Instituto de Astrof\'isica de Andaluc\'ia (IAA-CSIC), Glorieta de la Astronom\'ia s/n, E-18008 Granada, Spain
   \and
   Department of Physics and Astronomy, Chalmers University of Technology, Onsala Space Observatory SE-43992 Onsala, Sweden
   }

   \date{Received \dots; accepted \dots}

\abstract
{We present very long baseline interferometry (VLBI) observations of GRB\,221009A obtained with the global-VLBI network at 5\,GHz between 231 and 750\,days after the burst. The quality of our observations allowed us to measure the sub-milliarcsecond full-width at half maximum of the source surface brightness distribution with a one-sigma equivalent error of 13\% and 30\% at 231 and 391\,days post-burst, respectively. 
By fitting the temporal evolution of these new size measurements alongside those previously published in the literature at 5 and 8\,GHz with a single power law $\propto t^a$, we find a slope of $a = 0.52^{+0.14}_{-0.12}$ (median value with $1\sigma$ credible intervals). Using the public code \texttt{VegasAfterglow}, we model the 5\,GHz size evolution considering the afterglow emission of the forward shock produced by a jet decelerating in either a homogeneous interstellar medium (ISM) or into a surrounding medium with a wind-like density profile. The model constrains the collimation-corrected Sedov ratio, that is, the ratio of the jet kinetic energy $E_\mathrm{jet}$ to the density of the surrounding medium ($n$ in the ISM scenario, or $A_\star$ in the wind scenario), in the range $\log[(E_\mathrm{jet}/n)\,\mathrm{erg}^{-1}\,\mathrm{cm}^{-3}] = 54.2^{+0.4}_{-0.5}$ for the ISM scenario and $\log[(E_\mathrm{jet}/A_\star)\,\mathrm{erg}^{-1}] = 51.7^{+0.4}_{-0.2}$ for the wind scenario (median values with $1\sigma$ credible intervals). Among the works in the literature that fit the light curves and spectra of GRB\,221009A based on the standard afterglow model, and which can be reproduced with \texttt{VegasAfterglow}, none has best-fitting parameters that are compatible with these constraints within $2\sigma$. Moreover, we show that their predicted size evolution is not compatible with the observed one. Our results highlight the importance of size measurements of GRB afterglows as an independent probe of their dynamics, structure, and emission mechanisms.} 


\authorrunning{Giarratana, S. et al.}
\maketitle

\nolinenumbers

\section{Introduction}
GRB\,221009A is the brightest $\upgamma$-ray burst (GRB) detected to date \citep{lesage2023}. At a redshift of $z=0.151$, it also holds the record of the highest isotropic-equivalent energy ($E_\mathrm{\upgamma,iso} \gtrsim 10^{55}$\,erg). Its exceptionally bright afterglow was observed across the entire electromagnetic spectrum, from very high energy $\upgamma$-rays above 1\,TeV \citep{lhaaso2023, Cao2023}, down to the radio band, where it reached the highest flux density ever measured for a GRB \citep[approximately 60\,mJy at 17.7\,GHz;][]{Bright2023}. The unprecedented dataset collected has made GRB\,221009A one of the best-studied GRBs to date and provides a unique opportunity to probe the physics of relativistic blast waves. 

Despite the wealth of observations, the afterglow of GRB\,221009A has proven remarkably difficult to interpret within the framework of the standard afterglow model. In particular, the unusual evolution of the radio spectrum \citep{Bright2023, Laskar2023, Rhodes2024} cannot be explained by either a single forward shock (FS) or the combination of a FS and a reverse shock (RS) produced by a top-hat jet propagating into a cold external medium, with either a homogeneous (ISM) or a wind-like density profile \citep{Ren2023, Sato2023, Laskar2023, Kann2023, Zheng2024, Ren2024, Rhodes2024, Yao2025, Zhang2025, Barnard2025}. Although more complex scenarios were proposed to provide a better description of the data, such as a structured jet with an angular distribution in the energy and bulk Lorentz factor profiles \citep{OConnor2023, Gill2023, Sato2025}, the transition from a medium with a wind-like density profile to an ISM \citep{Ren2024, Zheng2024, Geng2025},  a Poynting-flux dominated jet \citep{Zhang2024} or time-evolving micro-physical parameters \citep{Foffano2024, Fraija2024}, no consensus on the physical properties of this afterglow has been reached yet. More generally, GRB\,221009A exposes a long-standing limitation of GRB afterglow modelling: the standard model is degenerate in the multi-dimensional parameter space (see, e.g., \citealt{Garcia-Cifuentes2024}). 
As a result, constraining the properties of the outflow and the environment surrounding the burst remains challenging, even for an event well sampled in the time and frequency space as GRB\,221009A. 

The milliarcsecond angular resolution of very long baseline interferometry (VLBI) radio observations allows direct measurements of the size of the emitting region, together with high-precision astrometry. For GRBs seen `on-axis', namely those with a viewing angle $\theta_{\mathrm{view}}$ between the observer line of sight and the GRB jet axis that is smaller than the jet half-opening angle $\theta_{\mathrm{jet}}$, the deceleration of the blast wave translates into the expansion of the projected image size. Conversely, if the outflow is observed `off-axis' ($\theta_{\mathrm{view}} > \theta_{\mathrm{jet}}$), the expansion of the shock wave primarily results into an apparent proper motion (see, e.g., \citealt{Granot2018}). The astrometric precision and angular resolution provided by VLBI observations have already been demonstrated to be fundamental to alleviate, or even break the degeneracy in the afterglow modelling \citep{Ghirlanda2019}. For the first merger of two neutron stars with a joint gravitational and electromagnetic detection, GW170817 / GRB\,170817A \citep{Abbott2017a, Abbott2017b, Margutti2021}, the light curves and spectra of the afterglow phase could be either explained by a jet with an angular structure of its energy and velocity distribution, or by an isotropic outflow with a distribution of its radial velocity profile \citep{Mooley2018a}. VLBI observations with the High Sensitivity Array and the global-VLBI network over 155 days revealed the apparent superluminal motion and constrained the size of the afterglow emission, providing strong support to the conclusion that the merger event resulted in the launch of a successful jet \citep{Mooley2018b, Ghirlanda2019, Mooley2022, Gourdji2026}.

In a previous work (\citealt{Giarratana2024}, hereafter \citetalias{Giarratana2024}), we reported the size expansion of GRB\,221009A, observed with the European VLBI Network (EVN) and the Very Long Baseline Array (VLBA) at 5, 8 and 15\,GHz. Our campaign provided evidence for the increase of the observed projected angular size of a GRB for the first time after GRB\,030329 \citep{Taylor2004, Taylor2005, Pihlstrom2007}. Moreover, the data hinted at the possible frequency dependent expansion due to different emitting components, with the 5 and 8\,GHz emission being dominated by the RS, while the FS contribution was predominant at 15\,GHz.
In this paper we present our new VLBI campaign of GRB\,221009A with the global-VLBI network, and we employ our measurements to test the models that were proposed in the literature to explain the multi-wavelength afterglow light curves. In Sec.\,\ref{sec:data}, we present the observations and the data calibration. Results are reported in Sec.\,\ref{sec:results}, while in Sec.\,\ref{sec:physical_model} we constrain the physical parameters of the afterglow through the modelling of the temporal evolution of the GRB size. In Sec.\,\ref{sec:discussion} we compare our size measurements with the predictions of the afterglow models proposed in the literature. Finally, we conclude with a summary in Section \ref{sec:conclusions}. 
Throughout the paper we assume a standard $\Lambda$--CDM model with $H_0 = 69.32$\,km\,Mpc$^{-1}$\,s$^{-1}$, $\Omega_m = 0.286$ and $\Omega_{\Lambda} = 0.714$ \citep{Hinshaw2013}. With these parameters, the luminosity distance at the redshift of GRB\,221009A ($z=0.151$) is $d_L = 726$\,Mpc, the angular diameter distance is $d_\mathrm{A} = 548$\,Mpc, and 1\,mas separation corresponds to 2.7\,pc in projection.

\section{VLBI observations and data reduction}
\label{sec:data}

\subsection{Observational setup}
We observed the field of GRB\,221009A with the global-VLBI network at 231 (2023 May 28), 391 (2023 November 4) and 750 (2024 October 28) days post-burst (PI: Giarratana, project codes: GG087, GG088). The global-VLBI network is a VLBI array comprising the EVN and the VLBA. These observations also included the participation of the southern hemisphere Long Baseline Array (LBA) and the phased-up Karl G. Jansky Very Large Array (VLA). The observations were performed at C-band (central frequency of 5\,GHz, total bandwidth of 512\,MHz). The data were recorded at bitrates of up to at 4\,Gbits\,s$^{-1}$. Dual polarisation products (RR, LL) were correlated at the Joint Institute for VLBI in Europe (JIVE, Dwingeloo, Netherlands) using the Super FX Correlator \citep[SFXC;][]{keimpema2015} into 16 sub-bands with a bandwidth of 32\,MHz and 64 channels each. 

The observations consisted of phase-referencing cycles with 4.5 minutes on GRB\,221009A and 1.5 minutes on the phase calibrator, namely the Very Large Array Sky Survey (VLASS) compact radio source J191142.50$+$195200 (J191142$+$1952 hereafter). The known VLBI calibrator J1925$+$2106 was employed as fringe finder. Throughout the observations, another standard  calibrator, J1905$+$1943, was also observed for 2\,minutes every 30--60\,minutes as `check' source, and treated like a target in the calibration procedure, in order to assess the quality of the phase-referencing. 

\subsection{Data calibration}

The data were calibrated using {\tt AIPS}\footnote{The Astronomical Image Processing System (AIPS) is a software package produced and maintained by the National Radio Astronomy Observatory (NRAO).} \citep{Greisen2003} and {\tt Difmap} \citep{Shepherd1994}, following the standard procedure for EVN phase-referenced observations\footnote{\url{https://www.evlbi.org/evn-data-reduction-guide}}. The amplitude calibration, which accounts for the antenna gain curves and the system temperature values measured by each telescope during the observation, was performed by applying the gains derived by the EVN pipeline. The calibration procedure was the following:
\begin{enumerate}
    \item we determined the single band delay with the \texttt{vlbampcl} procedure in {\tt AIPS} using multiple scans on J1925$+$2106, accounting for the fact that not all the antennas are observing the source at the same time in a global-VLBI observation;
    \item we performed the global fringe fitting on J191142$+$1952 and J1925$+$2106 with the \texttt{fring} task in {\tt AIPS}, without applying the band-pass correction computed by the EVN pipeline;
    \item to remove clear outliers, solutions from step 2 were then smoothed in time with the \texttt{snsmo} task in {\tt AIPS}, using \texttt{smotype}=`\texttt{dela}' and a time interval of 2\,hr. Solutions for rates and delays that deviate by more than 10\,mHz and/or 20\,nsec from their median values were removed. We then applied the solution table with the \texttt{clcal} task;
    \item we derived band-pass solutions with the \texttt{bpass} task using J1925$+$2106 and J191142$+$1952, both corrected for single-band delay and multi-band delay;
    \item we determined the single-band delay using the updated calibration table from step 3 and the band-pass solution in step 4. While, in principle, the band-pass acts only on the amplitudes, its correction is calculated with the complex visibility. As such, the band-pass can introduce some jumps in phase between different intermediate frequencies, affecting the results previously obtained with the single band delay. Therefore, we determined the single band delay again;
    \item we identified and flagged bad data in J191142$+$1952, and we propagated the flags to the GRB and the check source;
    \item to improve the quality of the calibration tables, steps 1 to 6 were repeated after flagging. This procedure was typically iterated 3-4 times, depending on the amount of flagging needed; 
    \item we split and exported the data of J191142$+$1952 with the \texttt{split} and \texttt{fittp} tasks in {\tt AIPS};
    \item we load the J191142$+$1952 data into {\tt Difmap} and produced a model of the source with the \texttt{modelfit} task;
    \item we performed a round of self-calibration of J191142$+$1952 in phase only with a solution interval of 30\,sec in {\tt Difmap}. We then inspected the solutions and the corrected visibilities for further flags. We repeated the self-calibration in phase with a solution interval of 10\,sec, followed by a final self-calibration in amplitude and phase with solution interval of 30\,sec. Depending on the epoch, multiple (up to two) faint components were detected, besides the bright core. Only those components robust under self-calibration were retained. Nevertheless, as their contribution in the total flux density was almost negligible compared to the core, their inclusion did not significantly affect the results; 
    \item we applied the flags observed in {\tt Difmap} to J191142$+$1952, the GRB and check source in {\tt AIPS};
    \item we performed the self-calibration of J191142$+$1952 in amplitude and phase in {\tt AIPS} using the image produced in {\tt Difmap} (step 9-10) as input;
    \item we applied the solutions to J191142$+$1952, the GRB and the check source in {\tt AIPS}. We split and exported the data with the \texttt{split} and \texttt{fittp} tasks in {\tt AIPS}. We then loaded the data into {\tt Difmap}, and we model-fitted and imaged the sources. To increase the signal-to-noise ratio, reduce the data size and the computational time, we averaged the visibilities over 60\,sec with the \texttt{uvaver} task in {\tt Difmap};
    \item to verify whether the phase-referencing was successful, we inspected the check source, J1905$+$1943, and flagged bad data. We found that the surface brightness peak and the size of the source are consistent within 1$\sigma$ confidence level in the first two observations. Conversely, the position of the centroid of the emission in right ascension disagrees by more than 5$\sigma$. Given that this inconsistency in the astrometry is most likely driven by the phase-referencing procedure, it prevents us from drawing any conclusions on the possible proper motion of the centroid of the GRB emission. We then propagated the flags to the GRB. This further step did not significantly change the results for the first two epochs, namely the size and flux density of the GRB before and after this step were consistent within 1$\sigma$. Concerning the third epoch, the absence of the Effelsberg antenna severely hampered the possibility to test the phase-referencing.
\end{enumerate}

\begin{table*}
\caption[]{Log table of our global-VLBI observations and results of the fits on the visibilities of the source with a circular Gaussian.}
\centering
\resizebox{\textwidth}{!}{
\begin{tabular}{llcrp{5cm}cccc|ccc}
\toprule
Code    &Date &Time  &$t_\mathrm{obs}-t_0$  &Antennas$^{\S}$ & $b_\mathrm{maj}^\dagger$ &$b_\mathrm{min}^\dagger$ &$b_\mathrm{p.a.}^\dagger$ &r.m.s.$^\dagger$ & $F_\nu^\star$     &  FWHM$^{\star}$ & $\langle \beta_\mathrm{app}\rangle^\star$           \\ 
& &[hh:mm UT]  &[days]  &  &[mas] &[mas] &[deg] &[$\upmu$Jy/b] &  [$\upmu$Jy]  &  [mas]         & c      \\ 
\midrule
GG087A &2023-05-28/29 &16:00 -- 16:00 &231 &At, Mp, Ho, Cd, T6, Jb, Wb, Ef, Mc, O8, Ys, Tr, Hh, Ib, Cm, Da, Kn, Pi, Yy, Br, Kp, Hn, Mk, Ov, Sc, Fd, La, Nl, Pt &3.88 &0.98 &$-8$ &2.5 &$180^{+10}_{-10}$ &$0.31^{+0.04}_{-0.04}$ &$2.4^{+0.3}_{-0.3}$\\
GG087B &2023-11-4/5 &05:30 -- 05:30 &391 &At, Mp, Hb, Ke, T6, Ur, Jb, Wb, Ef, Mc, O8, Tr, Hh, Ib, Sr, Cm, Da, De, Kn, Pi, Br, Kp, Hn, Mk, Ov, Fd, La, Nl, Pt, Yy &4.05 &2.29 &$-33$ &3.0 &$90^{+10}_{-10}$ &$0.68^{+0.12}_{-0.17}$ &$3.2^{+0.6}_{-0.8}$\\
GG088  &2024-10-28/29 &00:50 -- 06:30 &750 &Mp, Hb, Ke, T6, Jb, Wb, O8, Tr, Hh, Ir, Ys, Sr, Br, Kp, Hn, Mk, Ov, Sc, Fd, Nl, Pt, Yy &3.70 &1.52 &$-11$ &3.0 &$20^{+3}_{-3}$ &$<0.94$ &$<2.3$\\

\bottomrule
\end{tabular}}
\flushleft \footnotesize
$^{\S}$At = Australia Telescope Compact Array, Mp = Mopra 22\,m, Ho = Hobart 26\,m, Cd = Ceduna 30\,m, T6 = Tianma 65\,m, Jb = Jodrell Bank 76\,m, Wb = Westerbork 25\,m, Ef = Effelsberg 100\,m, Mc = Medicina 32\,m, O8 = Onsala 26\,m, Ys = Yebes 40\,m, Tr = Torun 32\,m, Hh = Hartebeesthoek 26\,m, Ib = Irbene 32\,m, Cm = Cambridge 32\,m, Da = Darnhall 25\,m, Kn = Knockin 25\,m, Pi = Pickmere 25\,m, Yy = phased-up VLA, Br = Brewster 25\,m, Kp = Kitt Peak 25\,m, Hn = Hancock 25\,m, Mk = Mauna Kea 25\,m, Ov = Owens Valley 25\,m, Sc = St. Croix 25\,m, Fd = Fort Davis 25\,m, La = Los Alamos 25\,m, Nl = North Liberty 25\,m, Pt = Pie Town 25\,m, Hb = Hobart 12\,m, Ke = Katherine 12\,m, Ur = Urumqi 25\,m, Sr = Sardinia Radio Telescope 65\,m, De = Defford 25\,m.\\
$^\dagger$Beam major axis, minor axis, position angle, r.m.s. noise level with natural weights.\\
$^\star$Median and 68\% confidence interval of the flux density $F_\nu$ and full width at half maximum FWHM from fitting a circular Gaussian to the source visibilities, and of the average apparent expansion speed $\langle \beta_\mathrm{app}\rangle$, assuming zero size at $t_0$. For GG088, we report the 95\% upper limit instead. The uncertainty on the flux density takes into account the statistical error on the fit.
\label{tab:log}
\end{table*}

\begin{figure*}
\centering
    \includegraphics[width=0.33\textwidth]{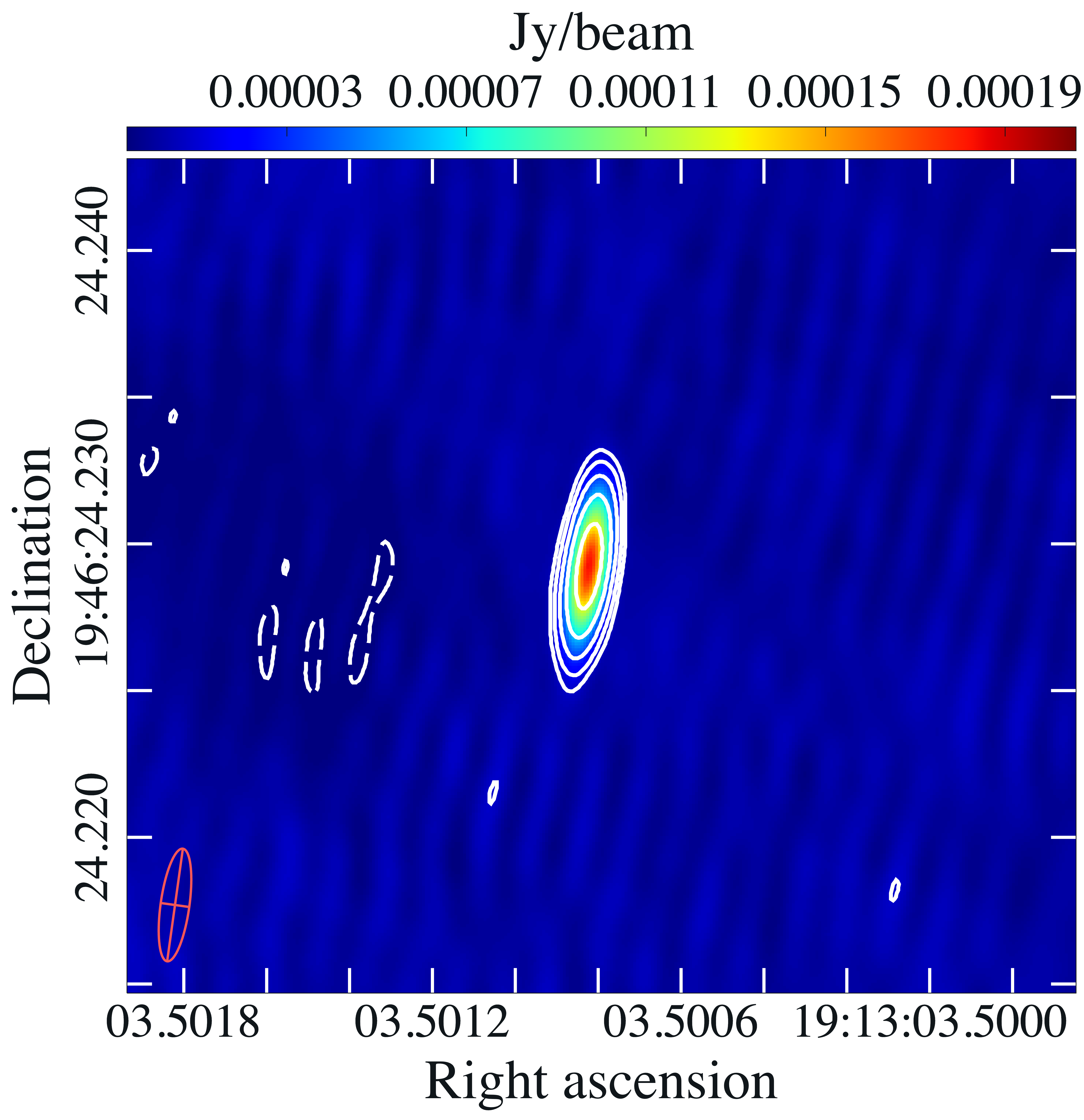}
    \includegraphics[width=0.33\textwidth]{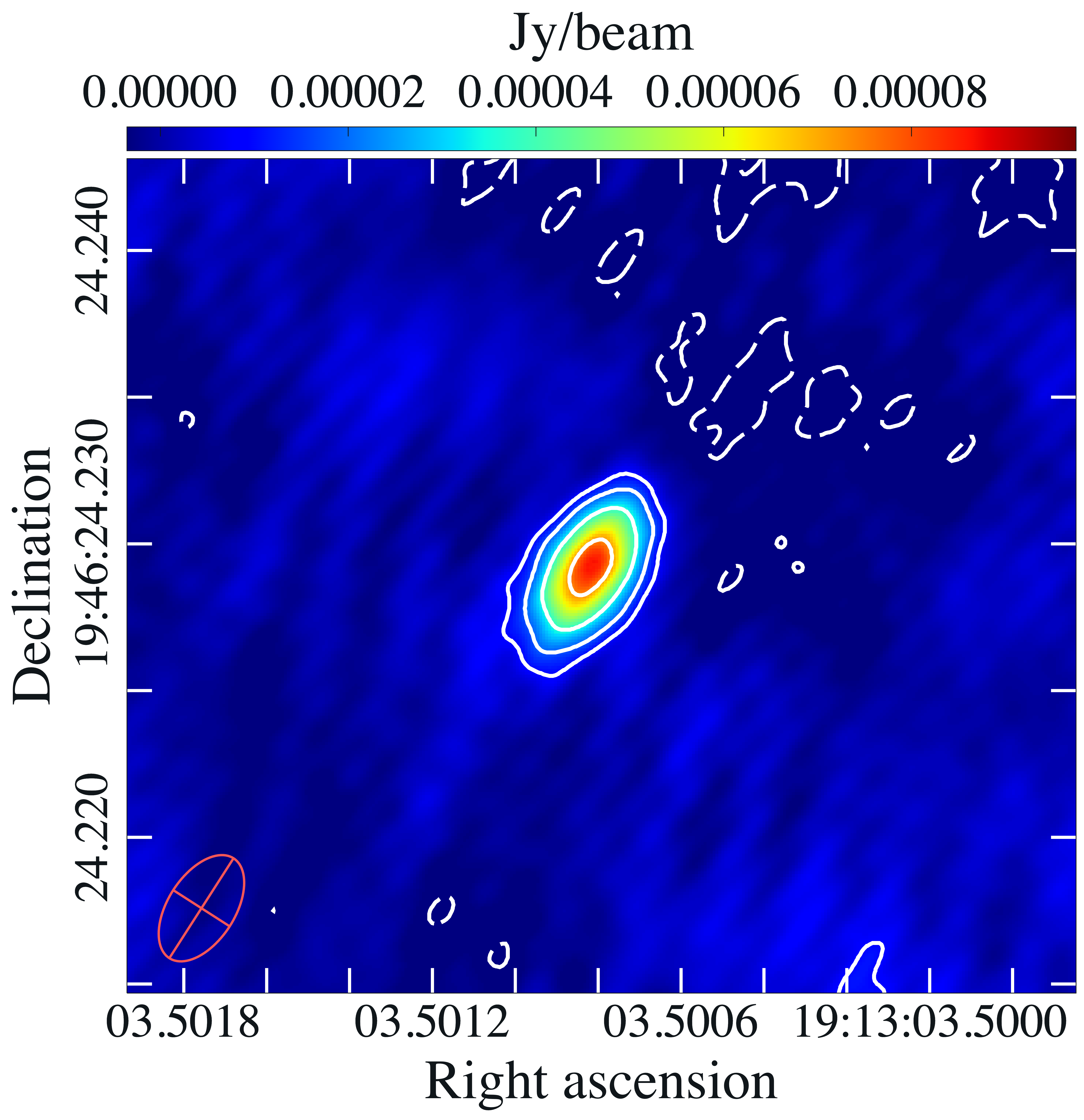}
    \includegraphics[width=0.33\textwidth]{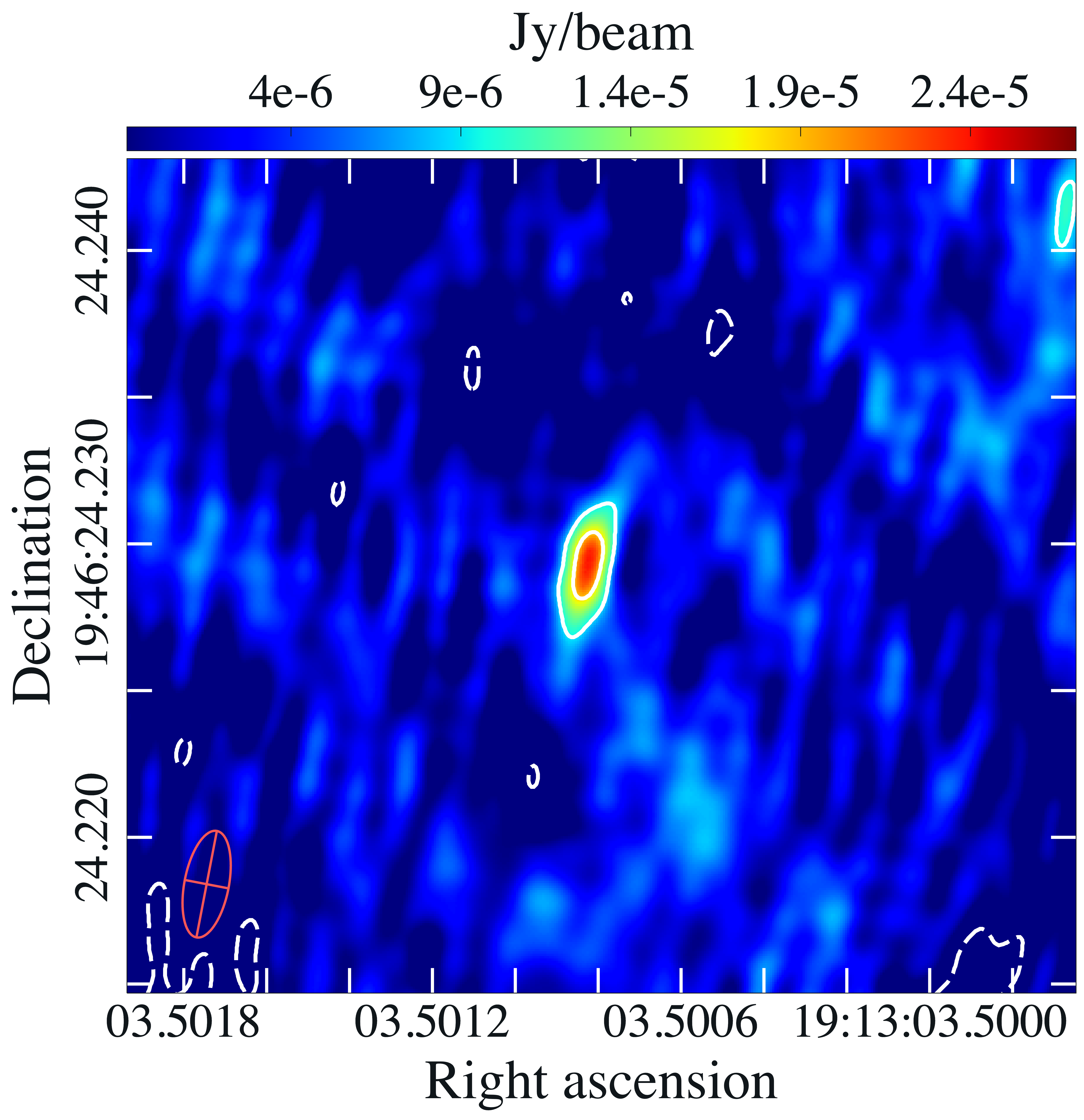}
    \caption{Radio maps of GRB\,221009A at 231 (left), 391 (centre) and 750 (right) days post-burst. Left panel: the contour levels at $3\sigma$, $6\sigma$, $12\sigma$, $24\sigma$, $48\sigma$ and $72\sigma$ are displayed as white solid lines, where $\sigma=2.5\,\upmu$Jy\,b$^{-1}$ is the r.m.s. noise level of the map, while the white dashed lines refer to the contour levels at $-3\sigma$. Central panel: the contour levels are at $-2\sigma$ (dashed lines), $3\sigma$, $6\sigma$, $12\sigma$ and $24\sigma$, where $\sigma=3\,\upmu$Jy\,b$^{-1}$. Right panel: the contour levels are at $-2\sigma$ (dashed lines), $3\sigma$ and $6\sigma$, with $\sigma=3\,\upmu$Jy\,b$^{-1}$. In each map, the de-convolved beam is displayed as a red ellipse on the bottom left. The rendering of the images was performed with \texttt{CARTA} (version 4.1.0; \citealt{CARTA2026}).}
    \label{fig:radio_maps}
\end{figure*}

The overall procedure yielded deep VLBI images of the GRB (see Fig.\,\ref{fig:radio_maps}), with r.m.s. noise levels of 2--3\,$\upmu$Jy\,beam$^{-1}$, depending on the epoch. Table\,\ref{tab:log} presents a summary of the properties of the observations: the code of the observation (first column), the date and time (UTC) of each epoch (second and third column), the mid time of the observations relative to the GRB trigger (fourth column), the antennas joining the observing array (fifth column), the de-convolved beam major and minor axes (sixth and seventh column) together with its position angle (eighth column), the r.m.s. noise level of the cleaned map (ninth column), the flux density (tenth column), the full
width at half maximum of the circular Gaussian fit performed on the visibilities (FWHM; eleventh column) and the computed apparent velocity (twelfth column, see Sec.\,\ref{sec:results}).


\section{Results}
\label{sec:results}

\subsection{Source brightness, size and average expansion speed}
Even for powerful and relatively close-by GRBs, the projected angular size of the afterglow image is typically smaller than the diffraction limit achieved in aperture synthesis interferometry with current facilities. Nevertheless, with sufficiently high signal-to-noise (S/N) observations, `super-resolution' can be achieved by fitting the visibilities of the source with a model \citep{Pearson1999, MartiVidal2012}. This allowed us to measure the size of the afterglow despite it being only a fraction of the beam size of our observations.

We estimated the total flux density, size and position of the afterglow following the method described in \citetalias{Giarratana2024}, which is a modified version of what was firstly presented by \cite{Salafia2022}: we fitted a circular Gaussian source model to the calibrated visibility data adopting a Markov Chain Monte Carlo (MCMC) approach, obtaining a posterior probability distribution of the flux density $F_{\nu}$, the FWHM of the circular Gaussian (proportional to the projected angular diameter of the source), and the position (RA and Dec) of the centre of the Gaussian. 

The first two global-VLBI observations had S/N values of approximately 70 and 30 at 231 and 391\,days post-burst, respectively. The $(u,v)$--plane coverage allowed us to measure the GRB angular size with relative uncertainties of approximately 13\% and 30\% at these two epochs (1-sigma equivalent, see Table\,\ref{tab:log}). The absence of the Effelsberg antenna during the final epoch at 750\,days post-burst, combined with the decreased source brightness, prevented a robust measurement of the source size. Therefore, for this epoch, we obtained only an upper limit on the angular size. We note that the flux densities, sizes and positions of the source as derived with our Bayesian approach are consistent with the results obtained with the \texttt{modelfit} task in {\tt Difmap}. Nevertheless, our method introduces amplitude correction factors to account for potential antenna-based systematic mis-calibrations (similar to \citealt{Natarajan2017}), allowing us to derive more conservative error bars (see \citetalias{Giarratana2024} for a full description).

\begin{figure}
\centering
    \includegraphics[width=0.4\textwidth]{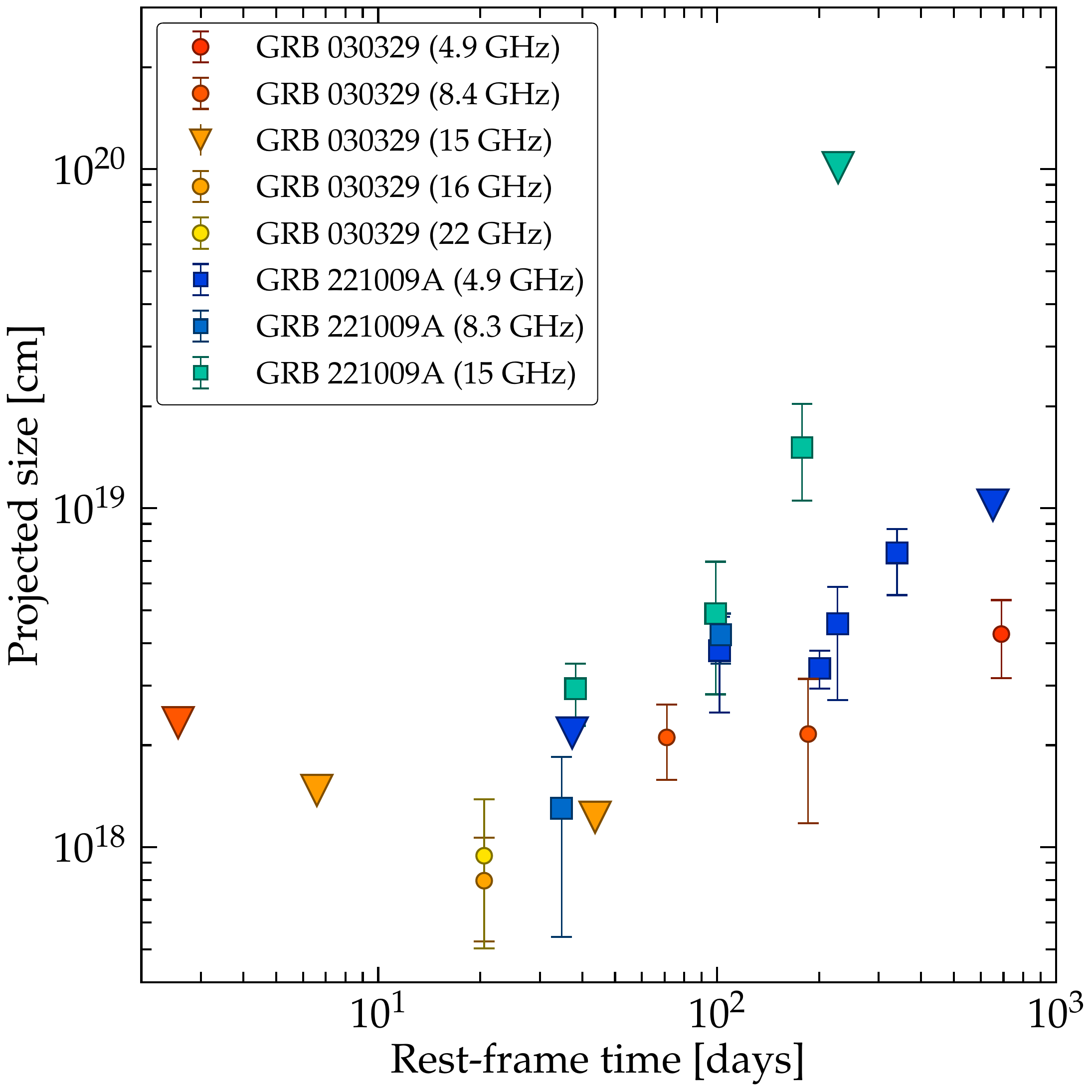}
    \includegraphics[width=0.4\textwidth]{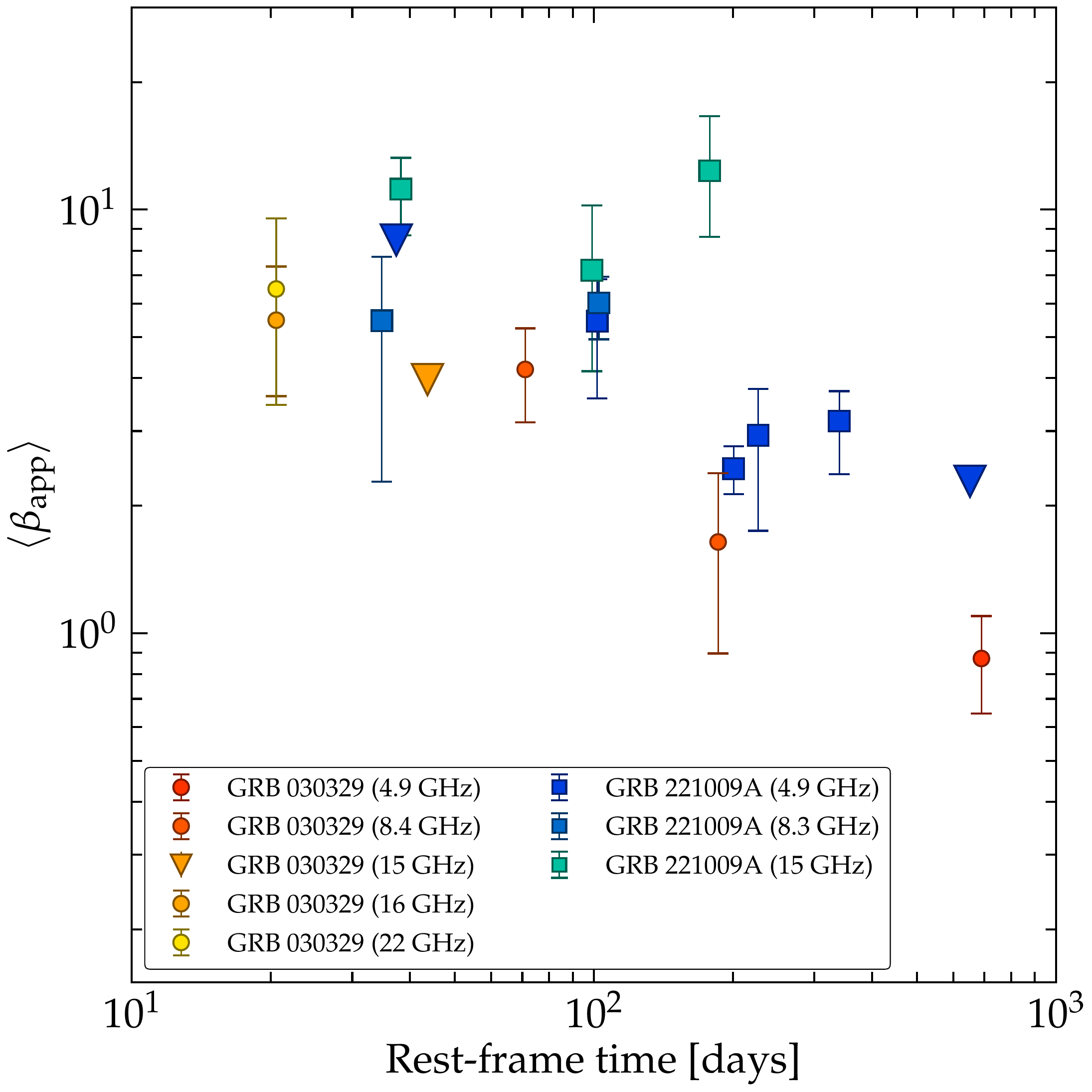}
    \caption{Temporal evolution of the projected size (upper panel) and the apparent expansion velocity (lower panel)  for GRB\,030329 \citep[circles;][]{Taylor2004, Taylor2005, Pihlstrom2007} and GRB\,221009A (squares; \citetalias{Giarratana2024} and this work), colour-coded according to the observing frequency. Upper limits are shown as triangles.}
    \label{fig:size_beta_vs_time}
\end{figure}

Assuming that $\mathrm{FWHM} = 0$ at the time of the explosion, the diameter can be used to calculate the average apparent expansion velocity $\langle\beta_{\mathrm{app}}\rangle$ as:
\begin{equation}
\label{eq:apparent_beta}
\langle\beta_{\mathrm{app}}\rangle = \frac{(1+z)d_\mathrm{A}}{2ct}\text{FWHM},
\end{equation}
where $t$ is the post-burst time of the observation, $c$ is the speed of light and $d_\mathrm{A}$ is the angular diameter distance. The average expansion velocities computed with Eq.\,\ref{eq:apparent_beta} are reported in Tab.\ \ref{tab:log}. The results for the first two epochs show a clearly superluminal expansion. In particular, our first global-VLBI observation at 231 days post-burst provides the most precise estimate of $\langle\beta_\mathrm{app}\rangle$ for a GRB to date. Figure \ref{fig:size_beta_vs_time} shows the temporal evolution of the projected size (upper panel) and the apparent expansion velocity $\langle\beta_\mathrm{app}\rangle$ (lower panel) for GRB\,030329 \citep{Taylor2004, Taylor2005, Pihlstrom2007} and GRB\,221009A (\citetalias{Giarratana2024} and this work). The evolution of the two quantities is broadly similar between the two GRBs, with GRB\,221009A showing some evidence for faster expansion and a larger image size at times $\gtrsim 300$\,d.

\subsection{Temporal slope of the expansion}

Assuming the afterglow emission at 5 and 8\,GHz to originate from the same emission mechanism, we fit the evolution of the angular size at these frequencies (including our data and those from \citetalias{Giarratana2024}) with a simple power-law model, following the same approach presented by \citetalias{Giarratana2024}:
\begin{equation}
s(t) = s_0 \biggr(\frac{t}{t_0} \biggl)^a,
\end{equation}
where $s_0$ is the size at the reference time $t_0 = 90$\,days post-burst. Given that the VLBA observations at 15\,GHz presented by \citetalias{Giarratana2024} may suffer from phase-de-correlation, and the nature of the emission they trace is still debated (see \citealt{Rhodes2024, Geng2025}), we did not include them in the fit. 

\begin{figure}
\centering
    \includegraphics[width=0.9\columnwidth]{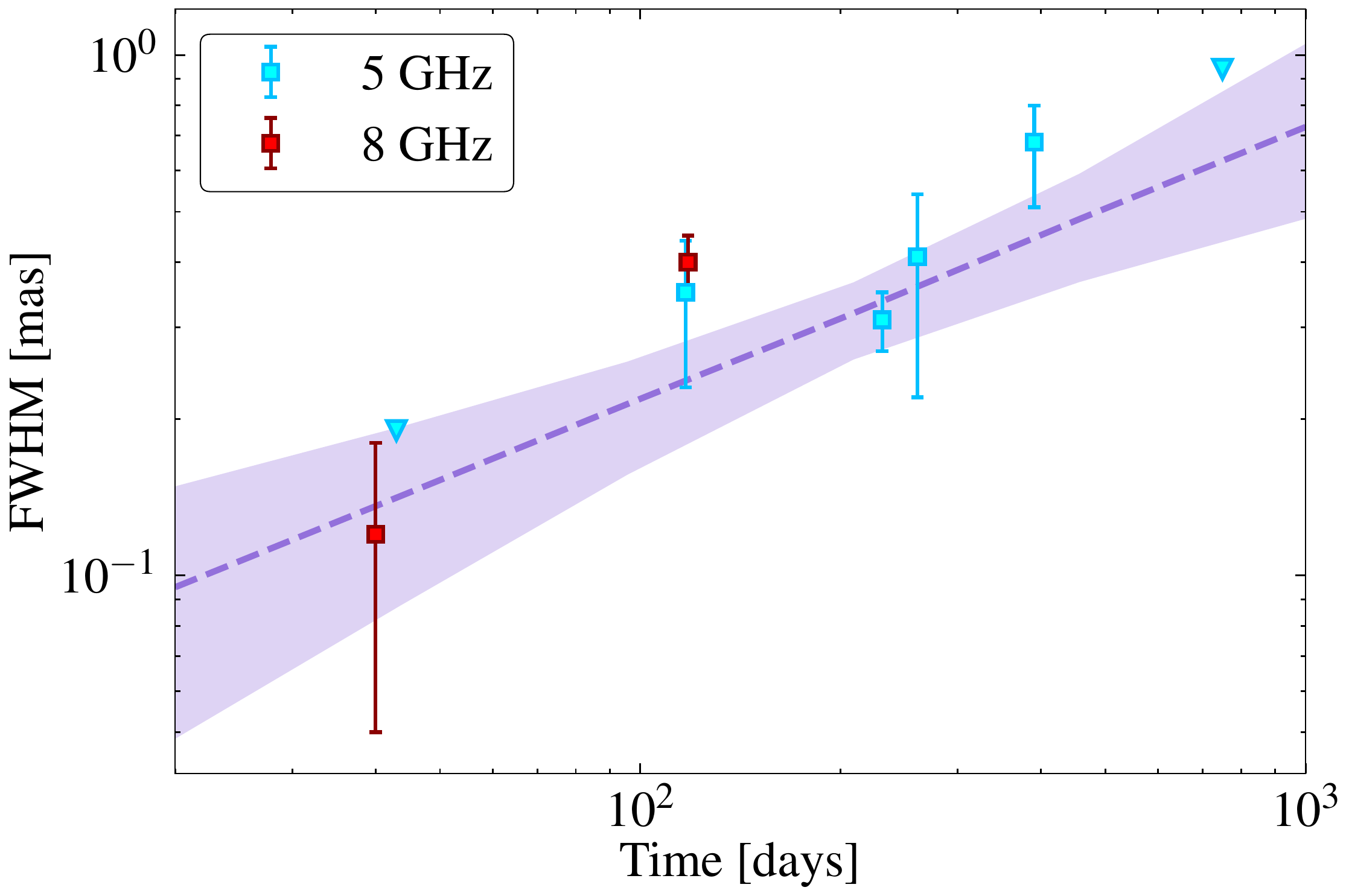}
    \includegraphics[width=0.9\columnwidth]{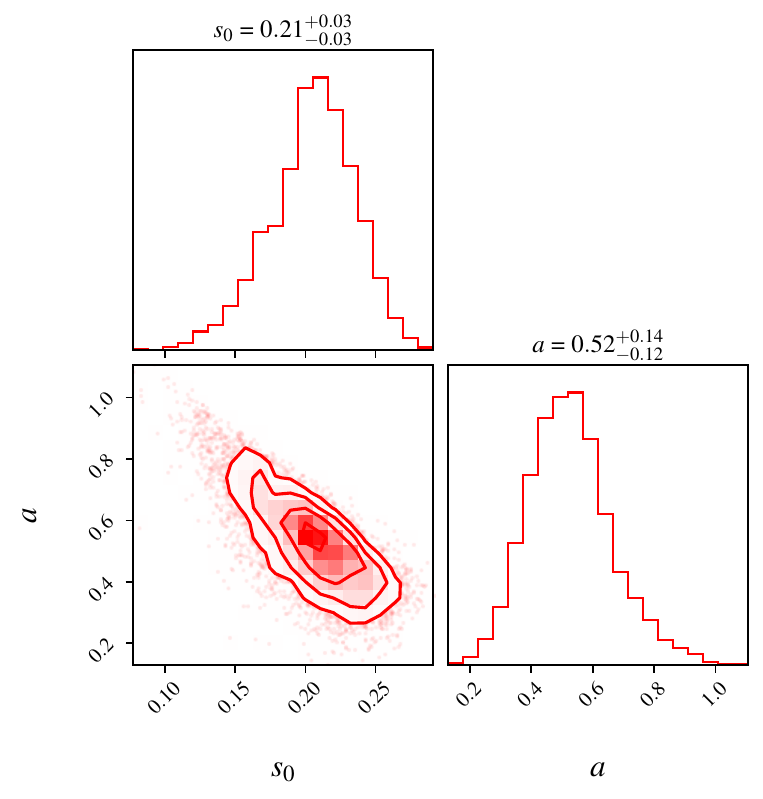}
    \caption{Size expansion (upper panel) and corner plot (lower panel) of the parameters obtained from the fit of the size measurements with a single power-law model. Upper panel: the cyan and red squares show all the observations (from both this work and \citetalias{Giarratana2024}) at 5 and 8\,GHz, respectively. The purple dashed line displays the median of the model realisations generated with the posterior distribution of the parameters $s_0$ and $a$, and the purple area presents the 90\% credible interval. Lower panel: the red histograms show the marginalised posterior probability densities.}
    \label{fig:power_law}
\end{figure}

Fig.\,\ref{fig:power_law} presents the size measurements of GRB\,221009A, together with the results from the fit. In the upper panel the observations at 5 and 8\,GHz are displayed as blue and red squares, respectively. The purple dashed line shows the median of the model posterior samples, whereas the purple area presents the 90\% credible interval at each fixed time. The lower panel of Fig.\,\ref{fig:power_law} shows the corner plot of the fit: red histograms represent the marginalised posterior probability densities, while the bottom-left panel shows one, two and three sigma equivalent contours of the joint posterior probability of $s_0$ and $a$.
We find  $a=0.52^{+0.14}_{-0.12}$ (median with 68\% credible interval). The posterior support for $a\leq0$ is $p(a\leq0)\sim 7\times 10^{-6}$, which implies that the source is expanding at a significance level of  approximately 4.5$\sigma$. 

The expansion temporal slope can be compared with the analytical predictions from  \cite{Granot2005}. In the ISM scenario, the expected slope before the time of the jet break $t_\mathrm{jet}$ is $a=5/8=0.625$. After this time, the slope transitions to $a=1/2=0.5$ in the case of exponential lateral expansion of the blast wave, or $a=1/4=0.25$ if the lateral expansion is negligible. Only the latter value is inconsistent with our inferred slope at a 1$\sigma$ confidence level. In the wind-like scenario, the expected slope is $a=3/4=0.75$ before the jet break and $a=1/2=0.5$ afterwards, regardless of the lateral expansion rate. In this case, the pre-jet-break value is not consistent with our inferred slope at a 1$\sigma$ confidence level. Therefore, although we are unable to distinguish between the ISM and wind-like scenarios, or between pre- and post-jet break, our data suggest a preference for a post-jet-break expansion and, in the case of an ISM, for a significant lateral expansion. 

\section{Physical modelling of the size expansion}
\label{sec:physical_model}

\subsection{Basic physical parameters of the size evolution}

To obtain more quantitative insight from the observed expansion, we compared the size predictions from a physical blast wave model with the observations. Before describing the model, we clarify the parameter space relevant to our problem.
The size evolution of the blast wave primarily depends on the Sedov length of the system, or equivalently on the ratio of the isotropic-equivalent kinetic energy $E_{\mathrm{iso}}$ and the number density of the external medium, which we indicate hereafter as the `Sedov ratio'. Defining the number density profile as $n(r) = Ar^{-k}$, where $r$ is the distance from the progenitor, we have $A=n$ and $k=0$ in the ISM case, or $A=3\times10^{35}A_{\star}$\,cm$^{-1}$ and $k=2$ in the wind case \citep{Panaitescu2000}, where 
\begin{equation}
A_{\star} = \frac{\dot{M}/(10^{-5}\,\mathrm{M_{\odot}\,yr}^{-1})}{v/(10^3\,\mathrm{km}\,\mathrm{s}^{-1})}    
\end{equation}
with $\dot{M}$ being the mass-loss rate of the progenitor massive star and $v$ the constant speed of its wind. Hence the Sedov ratio is $E_\mathrm{iso}/n$ in the ISM scenario, or $E_\mathrm{iso}/A_\star$ in the wind scenario. 

After $t_{\mathrm{jet}}$, the evolution depends also on the half-opening angle of the jet $\theta_{\mathrm{jet}}$ \citep{Granot2005}. The relation between $t_\mathrm{jet}$ and $\theta_\mathrm{jet}$ is \citep{Sari1999, Chevalier2000}:
\begin{equation}
\begin{split}
    t_{\mathrm{jet}} &\sim 1.5 (1+z) \left(\frac{\theta_{\mathrm{jet}}}{0.2\,\mathrm{rad}}\right)^{8/3}  \left(\frac{E_{\mathrm{iso,52}}}{n_{0}}\right)^{1/3}\,\mathrm{d} &\text{(ISM),}\\
    t_{\mathrm{jet}} &\sim 0.57(1+z) \left(\frac{\theta_{\mathrm{jet}}}{0.2\,\mathrm{rad}}\right)^{4}\left(\frac{E_{\mathrm{iso, 52}}}{A_{\star}}\right)\,\mathrm{d} &\text{(wind),}
\end{split}
\label{eq:thj_tj}
\end{equation}
where $E_{\mathrm{iso}, 52}$ is the isotropic-equivalent kinetic energy in units of $10^{52}$ erg. If most of the blast wave energy is contained within $\theta_\mathrm{jet}$, the expansion of the GRB afterglow image can be described by the combination of only two parameters: the Sedov ratio and $\theta_\mathrm{jet}$. If the angular structure is `shallow', with a significant amount of energy outside the `core', then the evolution may be more complex. Hereafter, we assume a `top-hat' structure for simplicity, namely a jet with uniform energy density within $\theta_\mathrm{jet}$, and zero outside, to derive physical constraints from the observed expansion in the simplest scenario.

\subsection{Size evolution model}

In order to map our two parameters to the size evolution, we employed the public code \texttt{VegasAfterglow}\footnote{\url{https://github.com/YihanWangAstro/VegasAfterglow}} \citep{VegasAfterglow2026, Zhang2018}, which incorporates advanced models for the afterglow evolution. \texttt{VegasAfterglow} computes the synchrotron emission from the electrons accelerated at the shock front, including self-absorption, and synchrotron self-Compton (SSC). It also allows to define complex jet structures and density profiles of the medium surrounding the burst. With \texttt{VegasAfterglow} it is possible to generate spatially resolved images (i.e.\ surface brightness maps) of the afterglow at any observer time and frequency with the \texttt{sky\_image} attribute of the \texttt{Model} class. Since the multi-wavelength light curves and spectra of GRB\,221009A exhibit a complex behaviour that is difficult to reproduce within the standard afterglow model (see e.g. \citealt{Laskar2023, Rhodes2024}), we restricted our fit to the sole evolution of the angular size. We defer an attempt to fit both the size evolution and the observed multi-wavelength light curves to a future work.

To simulate the synthetic radio sky map with \texttt{VegasAfterglow}, a set of physical and geometrical parameters must be defined in addition to the Sedov ratio and $\theta_\mathrm{jet}$. Since both $E_\mathrm{iso}$ and $n$ (or $A_\star$) must be defined independently, we fixed $n=1\,\mathrm{cm}^{-3}$ and $A_{\star}=1$ and derived $E_\mathrm{iso}$ based on the Sedov ratio. The other parameters needed are: the initial bulk Lorentz factor $\Gamma_0$, the fraction of the shock downstream internal energy retained by non-thermal electrons $\epsilon_{\mathrm{e}}$ and by the turbulent magnetic field $\epsilon_{\mathrm{B}}$, the power-law index $p$ of the distribution of electrons accelerated at the shock front, the viewing angle $\theta_\mathrm{view}$. For large values of the bulk Lorentz factor ($\Gamma_0\gtrsim100$), which are typically found in GRBs, the size does not vary significantly with $\Gamma_0$ \citep{Granot2005}: hence we fixed $\Gamma_0=100$. 
As the size evolution does not depend significantly on the microphysical parameters \citep{Granot2005}, we fixed them to typical values, namely $p=2.5$, $\epsilon_{\mathrm{e}}=0.1$ and $\epsilon_{\mathrm{B}}=10^{-3}$. 
We verified this assumption by simulating the size evolution expected from different combinations of the parameters and jet structures, varying $p$, $\epsilon_{\mathrm{e}}$ and $\epsilon_{\mathrm{B}}$ while leaving the other parameters fixed. The allowed ranges were $[2.01,2.9]$, $[10^{-3}, 0.5]$ and $[10^{-6}, 0.5]$ for $p$, $\epsilon_{\mathrm{e}}$ and $\epsilon_{\mathrm{B}}$, respectively. We also fixed $\theta_\mathrm{view}=0$\,deg, which reflects the likely small viewing angle of the GRB \citep{OConnor2023} and the fact that the viewing angle is most likely unimportant at the late times of our observations. We found that changing the micro-physical parameters resulted in a negligible difference of only a few percent in the expected size. 

We adopted the wide priors $E_\mathrm{iso}/n \in [10^{54}, 10^{59}]$\,erg\,cm$^{-3}$ for the ISM and  $E_\mathrm{iso}/A_{\star} \in [10^{52}, 10^{56}]$\,erg for the wind-scenario. Given the uncertainty on the presence of a jet break signature in the light curve of GRB\,221009A \citep{lhaaso2023, Levan2023, Rhodes2024, Sears2025}, we set the prior on $\theta_\mathrm{jet}$ such that $t_{\mathrm{jet}}\in[10^{-3},10^{3}]$\,days post-burst. 

We included the effects of lateral expansion with the \texttt{spreading} parameter in the \texttt{Model} class of \texttt{VegasAfterglow}. Since the radio emission of GRB\,221009A after 100\,days post-burst is believed to be dominated by the FS \citep{OConnor2023, Gill2023}, we did not include the contribution from the RS, and we limited the fit to the measurements performed after this date. 
For a given set of parameters, we produced a synthetic radio image at each time of the observations, represented by a set of surface brightness values $I_{i,j}$ corresponding to the indices $(i,j)$ of the pixels in the image. 

\subsection{Comparison with the observations}
Because of the small size of the source, our observations only probe the first-order Taylor expansion of the Fourier transform of the surface brightness \citep{Interferometry_book}. In such a situation, the FWHM of our circular Gaussian model can be directly compared to the first moment of the surface brightness distribution, which we computed as
\begin{equation}
\label{eq:fwhm}
\mathrm{FWHM} = \sqrt{8\ln 2}\left(\frac{\iint I  R^2\, dX\,dY}{2\iint I \, dX\,dY}\right)^{1/2}\approx2.355\sqrt{\frac{\sum_{i,j}\biggr(I_{i,j} R_{i,j}^2\biggl)}{2\sum_{i,j} I_{i,j}}},
\end{equation}
where $X$ and $Y$ are the angular coordinates of the image, $R=\sqrt{X^2+Y^2}$ is the angular distance from the centre of the source image, and the last approximation is valid for an image with fixed pixel size. This definition yields the correct FWHM in case of a circular Gaussian, and we argue that it correctly represents the FWHM of the best-fitting circular Gaussian model of a more general surface brightness distribution, in the case where the source image is smaller than the beam (i.e.\ the source is `super-resolved'; \citealt{MartiVidal2012}). We tested this by generating several model images with different parameters, producing synthetic visibilities from these images with the \texttt{eht-imaging} package \citep{Chael2018,Chael2019}, and fitting a circular Gaussian model to these with \texttt{Difmap}. The FWHM of the best-fitting circular Gaussian always matched within 10-20\% the value computed directly from the image with the above formula (see Appendix \ref{appendix:check_FWHM} for more details).

This correspondence allowed us to associate a FWHM to our physical model of the size evolution and hence fit it to the VLBI observations at 5\,GHz (including our data and those from \citetalias{Giarratana2024}) within a Bayesian framework, sampling the posterior probability distributions with the MCMC algorithm implemented in the \texttt{emcee} python package \citep{Foreman-Mackey2013}. We ran 5000 iterations of the MCMC with 50 walkers for each scenario (ISM or wind). 

\begin{figure*}
\centering
    \includegraphics[width=0.95\columnwidth]{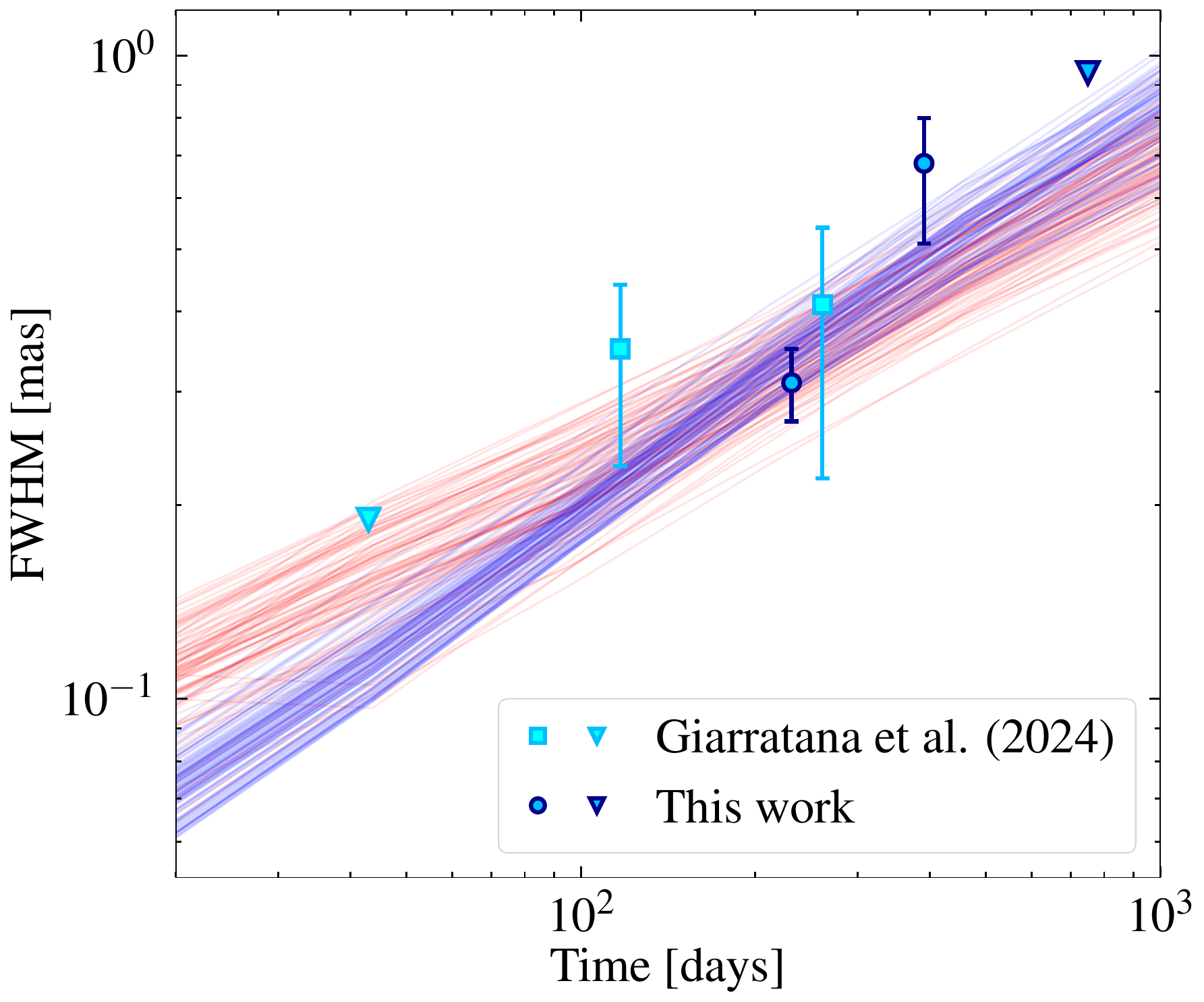}\vspace{5pt}\includegraphics[width=0.95\columnwidth]{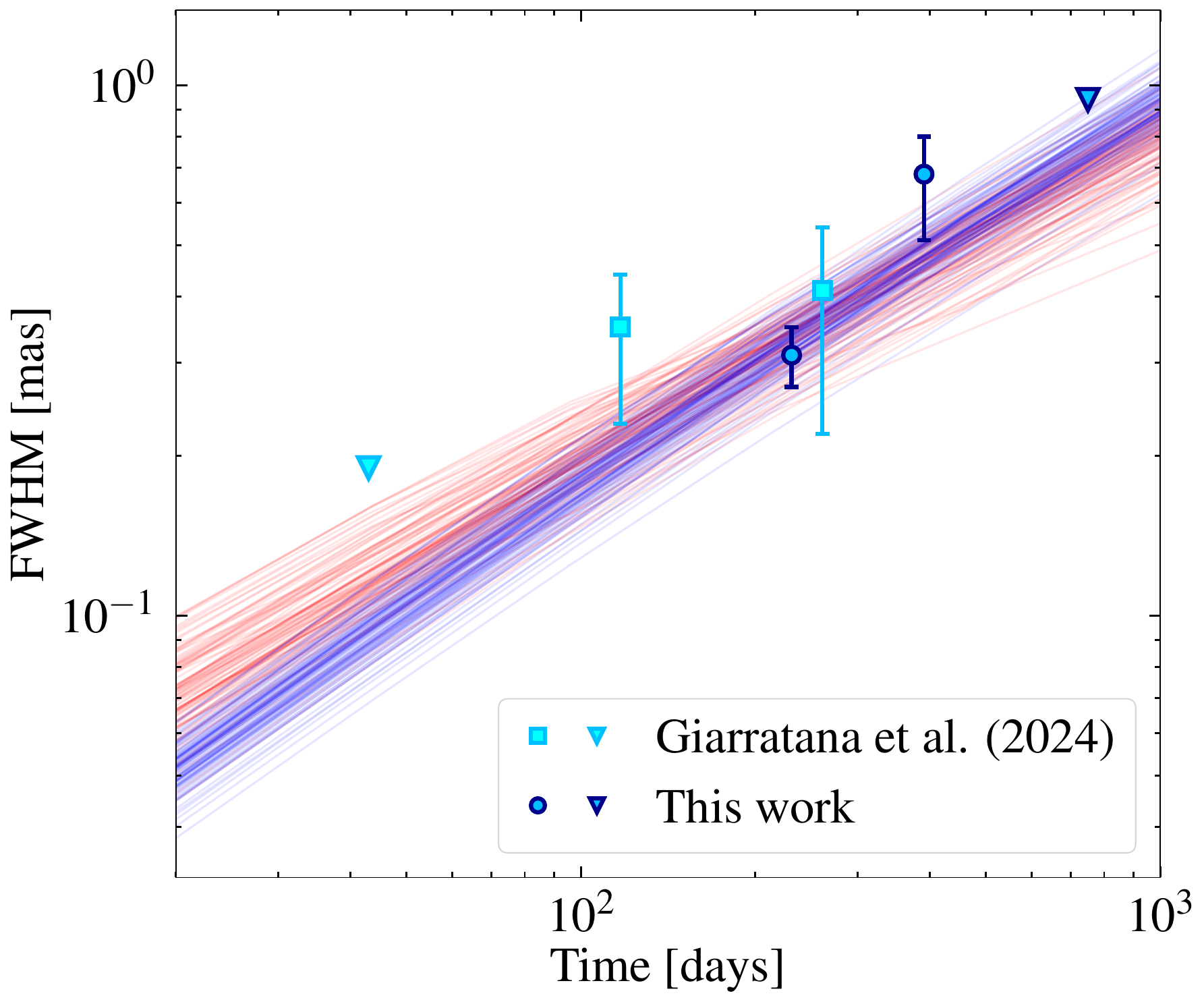}\\
    \includegraphics[width=0.95\columnwidth]{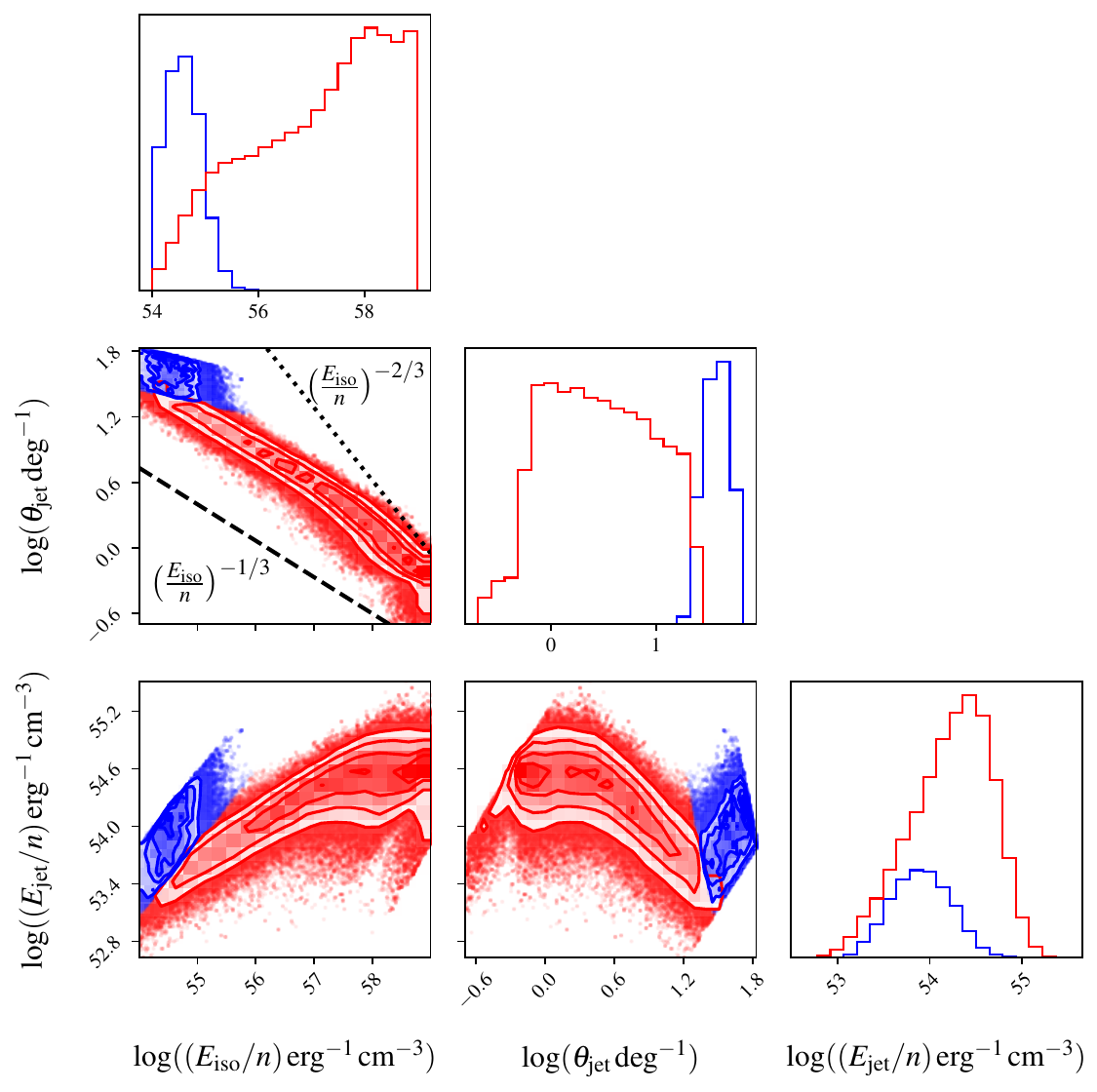}\vspace{5pt}\includegraphics[width=0.95\columnwidth]{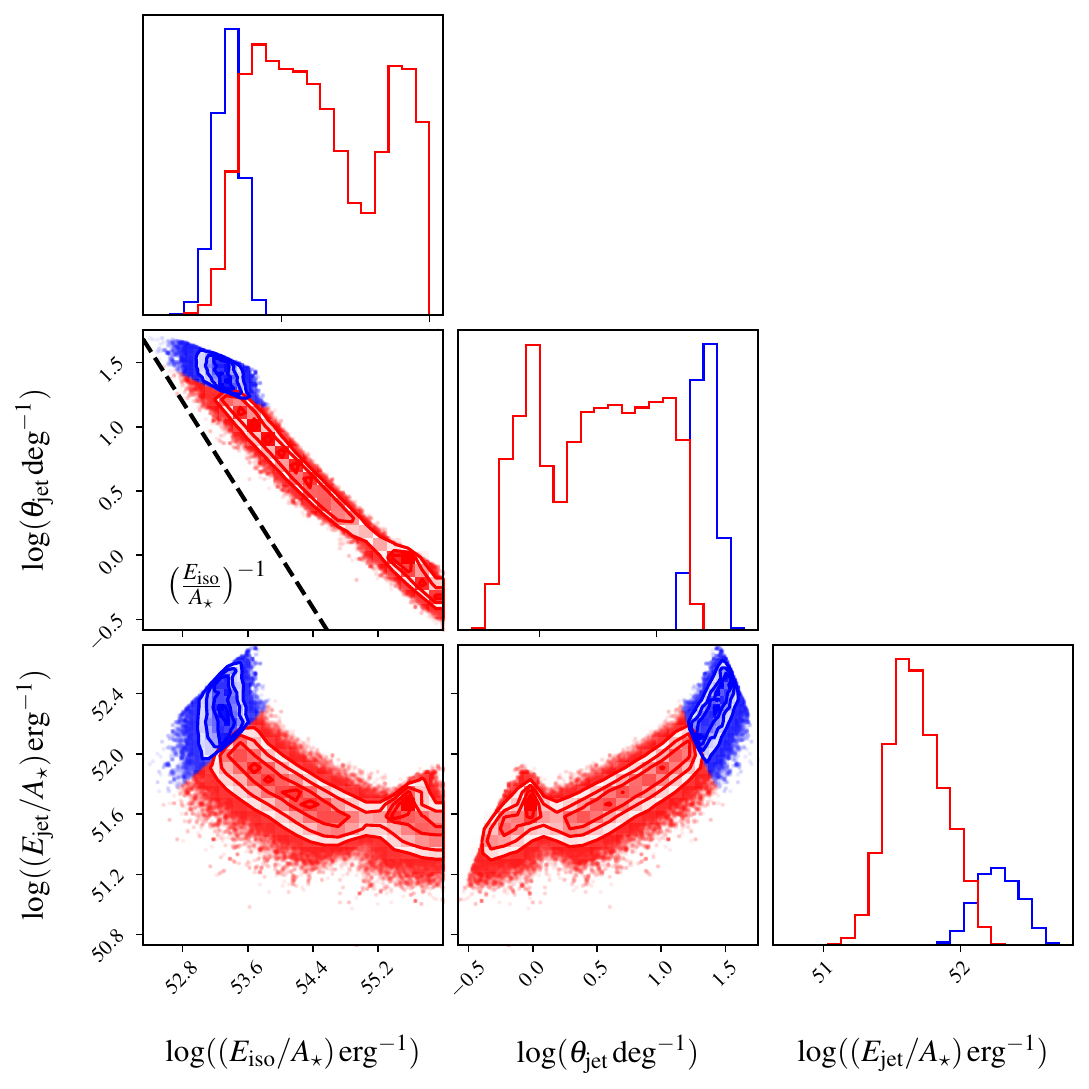}\\
    \caption{Size expansion at 5\,GHz (upper panels) and corner plots (lower panels) of the parameters obtained from the fit of the physical model to our size measurements in the ISM scenario (left-hand panels) and the wind scenario (right-hand panels). The solid lines in the upper panels show size evolution models generated with 100 random posterior samples of the fit for $t_\mathrm{jet}>200$ days (blue) and $t_\mathrm{jet}<200$ days (red). The lower panels show corner plots of the posterior probability densities constructed from the subset of posterior samples with $t_\mathrm{jet}>200$ days (blue) and  $t_\mathrm{jet}<200$ days (red). The analytical relation between the Sedov ratio and the half-opening angle of the jet from \citet{Granot2005} are shown as dashed and dotted lines (see the text for the explanation).}
    \label{fig:ism+wind}
\end{figure*}

Fig.\,\ref{fig:ism+wind} summarizes the results of our fits. The upper panel of each figure shows 100 model realisations obtained by randomly sampling the Sedov ratio and $\theta_\mathrm{jet}$ from the posterior distribution. The lower panel displays the corresponding corner plot. While we generated a single posterior distribution, we coloured in red and blue the subsets of the parameter space for which the jet break predicted by \texttt{VegasAfterglow}\footnote{We find that the jet break time in \texttt{VegasAfterglow} occurs later than predicted by Eqs.\ \ref{eq:thj_tj} by around a factor of two, but we note that the transition is not sharp, hence this value is intrinsically approximate.} occurs before and after approximately 200\,days post-burst, respectively. This distinction was introduced because the size evolution before $t_\mathrm{jet}$ does not depend on $\theta_\mathrm{jet}$, and this is reflected by the fact that the joint posterior becomes vertical in the Sedov ratio \textit{vs} half-opening angle plane (i.e., independent of $\theta_\mathrm{jet}$) if the observations are before $t_\mathrm{jet}$. 

We found that $\log[(E_\mathrm{iso}/n)\,\mathrm{erg}^{-1}\,\mathrm{cm}^{-3}] = 56.7^{+1.6}_{-2.1}$ and $\log[(E_\mathrm{iso}/A_{\star})\,\mathrm{erg}^{-1}] = 54.2^{+1.4}_{-0.8}$ (median values of the entire distribution with $1\sigma$ credible intervals) in the ISM and wind-like scenario, respectively. We included $\log(E_\mathrm{jet}/n)$ or $\log(E_\mathrm{jet}/A_\star)$ in the corner plot, where $E_\mathrm{jet}=E_\mathrm{iso}(1-\cos\theta_\mathrm{jet})$ is the collimation-corrected energy. We found that $\log[(E_\mathrm{jet}/n)\,\mathrm{erg}^{-1}\,\mathrm{cm}^{-3}] = 54.2^{+0.4}_{-0.5}$ and $\log[(E_\mathrm{jet}/A_{\star})\,\mathrm{erg}^{-1}] = 51.7^{+0.4}_{-0.2}$ (median values of the entire distribution with $1\sigma$ credible intervals) in the ISM and wind-like scenario, respectively. In the ISM scenario, the result implies a very large jet total energy, unless $n$ is orders of magnitude below the standard value $n=1\,\mathrm{cm^{-3}}$. By contrast, the wind scenario can accommodate a more reasonable jet total energy in the case of a standard wind parameter.

Our first global VLBI epoch (GG087A) provides the tightest constraint to date on the size of the GRB\,221009A afterglow. This forces any model to reproduce that specific size at the time of that observation. In the post-jet break scenario, our constraint translates to a specific relation between the Sedov ratio and the half-opening angle in the analytical models of \citet{Granot2005}, which depends on the lateral expansion rate in the ISM case. We show these induced relations in the corner plots of Fig.\,\ref{fig:ism+wind}, for comparison. In the ISM scenario, $E_\mathrm{iso}/n \propto \theta_\mathrm{jet}^{-1/3}$ if the lateral expansion of the jet is negligible (black dashed line in the corner plot on the bottom left), while $E_\mathrm{iso}/A_{\star} \propto \theta_\mathrm{jet}^{-2/3}$ in the case of exponential lateral expansion (dotted line in  in the corner plot on the bottom left). As expected, the posterior distribution of our fit follows the exponential lateral expansion relation if the initial half-opening angle is small, and the relation for negligible lateral expansion otherwise, with a transition at around $\theta_\mathrm{jet}\sim 3$ deg. In the wind scenario, the model by \citet{Granot2005} predicts $E_\mathrm{iso}/A_{\star} \propto \theta_\mathrm{jet}^{-1}$, independently of the effects of lateral expansion (black dashed line in  in the corner plot on the bottom right). The posterior distribution follows a similar trend for relatively large half-opening angles, while the slope is different for small half-opening angles. This likely reflects the fact that the analytical model does not fully capture the effects of lateral expansion in the wind scenario.

\section{Discussion}
\label{sec:discussion}

In what follows, we compare our results for GRB\,221009A with our previous work \citepalias{Giarratana2024}, with GRB\,030329, and with studies that model the light curves and spectra of the afterglow of this GRB, namely \citet{Ren2023}, \citet{Sato2023}, \citet{OConnor2023}, \citet{Gill2023} and \citet{Ren2024}. 
We considered also some additional models \citep{Laskar2023,Zheng2024,Foffano2024,Fraija2024,Yao2025,Zhang2025,Sato2025,Geng2025}, but different reasons prevented a direct comparison with these, as we explain below.

\subsection{Collimation-corrected Sedov ratio}

\begin{figure}
\centering
\includegraphics[width=0.9\columnwidth]{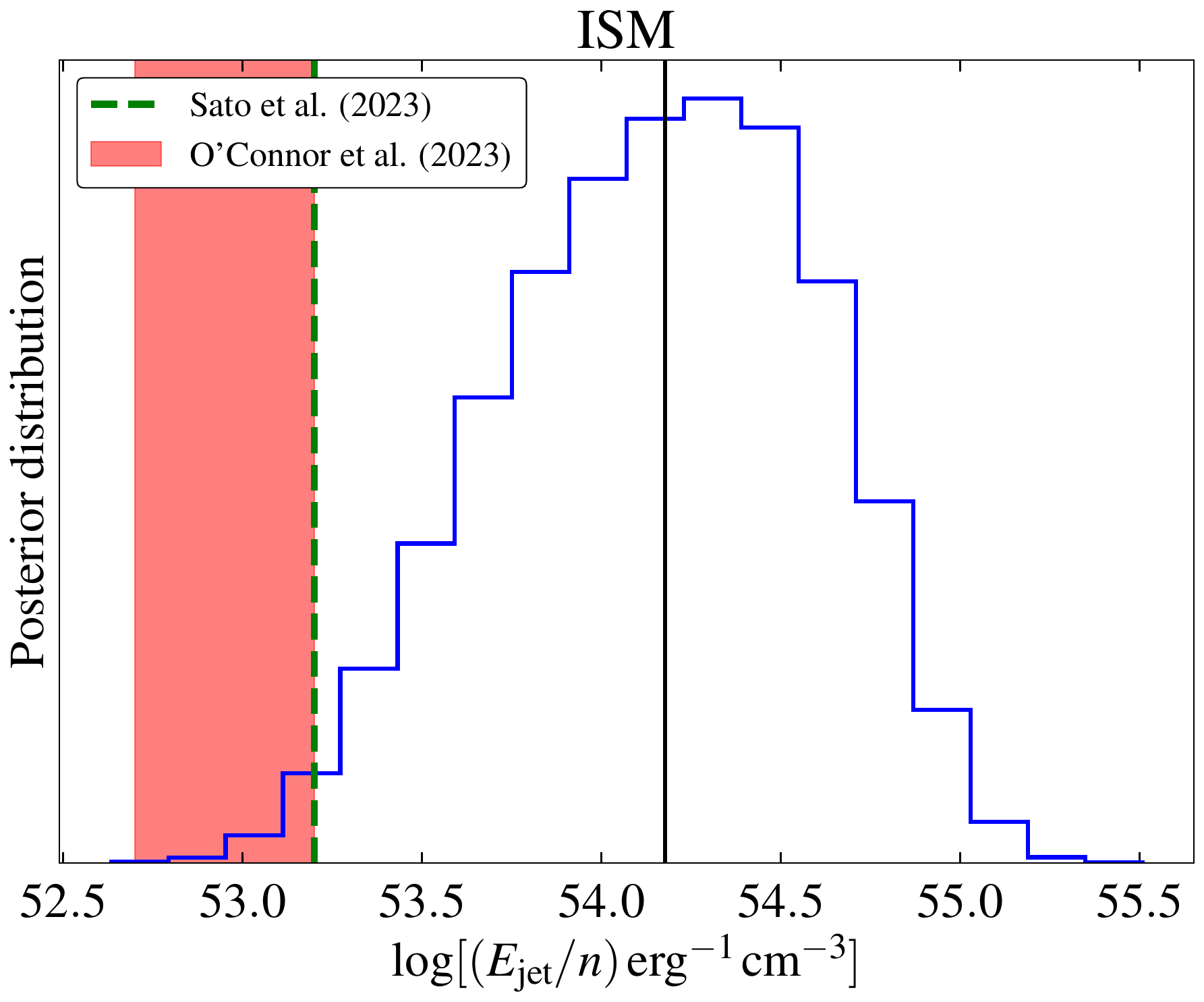}
\includegraphics[width=0.9\columnwidth]{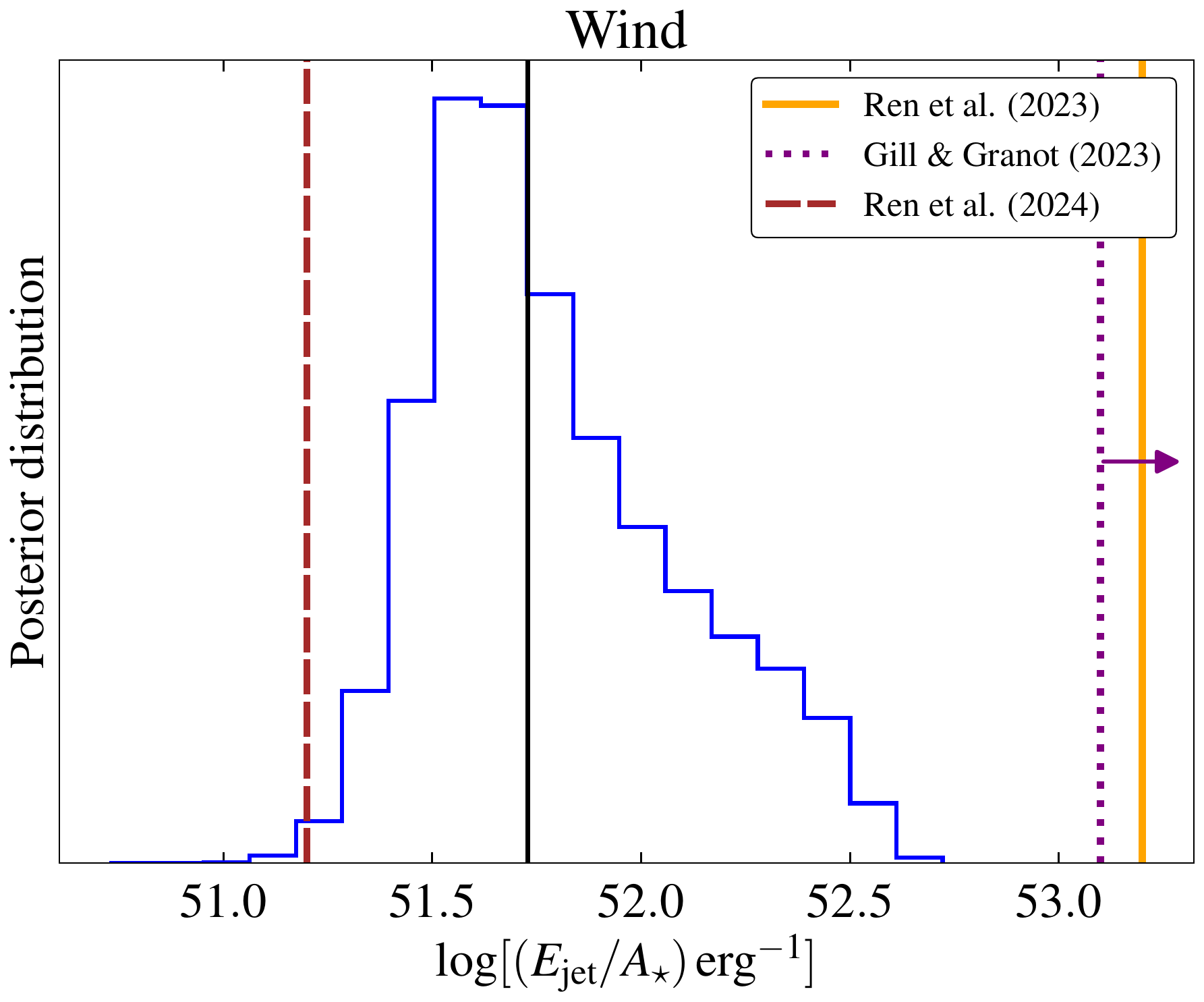}
    \caption{Posterior distribution of the logarithm of $E_\mathrm{jet}/n$ in the ISM scenario (upper panel) and of $E_\mathrm{jet}/A_\star$ in the wind (lower panel) scenario. The coloured vertical lines refer to the collimation-corrected Sedov ratio inferred from the best fit parameters of each proposed model. Upper panel: \cite{Sato2023} (green dashed line), \cite{OConnor2023} (red area). Lower panel: \cite{Ren2023} (orange solid line), \cite{Gill2023} (purple dotted line) and \cite{Ren2024} (brown long-dashed line). The arrow close to the vertical dotted line indicates that the allowed range of the \citet{Gill2023} model extends to the right of the plot. In each panel, the black vertical line pinpoints the median of the distribution.}
\label{fig:Ejet_hist}
\end{figure}

\begin{table}
    \caption{Collimation-corrected Sedov ratios.}
    \small
    \centering
    \begin{tabular}{lccc}
    \toprule
    Reference & k & $\log\left(\frac{E_\mathrm{jet}/n}{\mathrm{erg\,cm^{3}}}\right)$ & $\log\left(\frac{E_\mathrm{jet}/A_\star}{\mathrm{erg}}\right)$\\
    \midrule
    This work       & 0 & $54.2^{+0.4}_{-0.5}$ & ~\\
    This work       & 2 & ~ & $51.7^{+0.4}_{-0.2}$\\
    \citetalias{Giarratana2024} & 0 & $55.8\pm0.5$ & ~\\
    \citetalias{Giarratana2024} & 2 & ~ & $53.8\pm0.3$\\
    \citet{Ren2023} & 2 & ~ & $53.2$ \\
    \citet{Sato2023} & 0 & $53.2$ & ~ \\
    \citet{OConnor2023} & 0 & $52.7-53.2$ & ~ \\
    \citet{Gill2023} & 2 & ~ & $53.1-54.5$ \\
    \citet{Ren2024} & 2$^\dagger$ & ~ & $51.2$ \\
    \bottomrule
    \end{tabular}
    \label{tab:Sedov}
    \tablefoot{$^\dagger$The circum-burst medium density profile of \citet{Ren2024} is non-standard, but it is wind-like at the large radii relevant to this study.}
\end{table}

The collimation-corrected Sedov ratios $E_\mathrm{jet}/n$ and $E_\mathrm{jet}/A_\star$ inferred from our fits are relatively well-constrained, with only about one order of magnitude of uncertainty. The constraints are not consistent with those reported by \citetalias{Giarratana2024}: using their posterior samples, we obtain $\log[(E_\mathrm{jet}/n)\,\mathrm{erg}^{-1}\,\mathrm{cm}^{-3}] = 55.8\pm 0.5$ and $\log[(E_\mathrm{jet}/A_{\star})\,\mathrm{erg}^{-1}] = 53.8\pm0.3$, both of which are in tension at $>2\sigma$ with the results from the present study. However, we note that in \citetalias{Giarratana2024} the emission at 5\,GHz was assumed to originate from the RS, whereas in this work we attribute the emission beyond 100\,days post-burst to the FS, in agreement with more recent models presented in the literature (see e.g. \citealt{OConnor2023, Gill2023}). Hence, the assumed interpretation of the source whose image size is measured influences the conclusions on the physical parameters of the explosion.

In Table \ref{tab:Sedov} and Figure \ref{fig:Ejet_hist}, we compare our collimation-corrected Sedov ratio posterior distributions with the best fit values from the models from the literature mentioned above. Since most of these models assume structured jets, we define their collimation-corrected Sedov ratio using the total jet energy that is obtained by integration of their assumed jet energy structure. The outer half-opening angle in the models  of \citet{OConnor2023} and \citet{Gill2023} is only defined as a lower limit due to the absence of an observed jet break, and their structures are shallow, so that the energy content is sensitive to this angle. Therefore, for these models we report the range span by the collimation-corrected Sedov ratio when the outer half-opening angle is comprised between the minimum value quoted in the original article and $\pi/2$. The model of \citet{Ren2024} assumes a non-standard circum-burst medium. However, given that their density profile is wind-like from a transition radius $r_\mathrm{tr}\approx 3.7\times 10^{16}\,\mathrm{cm}$ outwards, and since this radius is long surpassed at the times relevant for our study, we consider a wind-like scenario for this work. Inspection of Figure \ref{fig:Ejet_hist} shows that all considered models, except those from \citet{OConnor2023} and \citet{Sato2023}, predict collimation-corrected Sedov ratios that are incompatible at $\gtrsim2\sigma$ with our results.

\subsubsection{Comparison with GRB\,030329}

It is instructive to compare the collimation-corrected Sedov ratio of GRB\,221009A with that of GRB\,030329, the only other GRB whose size and expansion rate have been constrained through VLBI observations. Studies of the latter GRB that combine image size and spectral information \citep{Granot2005,vanderHorst2008,Mesler2013} find $\log[(E_\mathrm{jet}/n)\,\mathrm{erg^{-1}\,cm^{-2}}]$ in the range 50.5 -- 51.7 when assuming an ISM-like medium, or  $\log[(E_\mathrm{jet}/A_\star)\,\mathrm{erg^{-1}}]$ in the range 51.1 -- 51.7 when assuming a wind-like medium. However, we note that these studies 
relied on a different relation between measured and physical source size, $\mathrm{FWHM}\sim 2R_\perp/d_\mathrm{A}$, where $R_\perp$ is the projected radius of the source image \citep{Granot1999}. At fixed $\mathrm{FWHM}$, such relation underestimates $R_\perp$ typically by a factor of about 0.65 \citep{Salafia2022}, which translates into a stronger underestimation of the Sedov ratio that depends on the expansion regime and density profile, being more severe for the ISM than for the wind scenario. Therefore, in order to obtain an estimate of the collimation-corrected Sedov ratio of GRB\,030329 that is more directly comparable with that derived here for GRB\,221009A, we fitted the image size evolution of GRB\,030329 with the approach described in Sec.\,\ref{sec:physical_model}. We find $\log[(E_\mathrm{jet}/n)\,\mathrm{erg^{-1}\,cm^{-2}}]=52.9^{+0.4}_{-0.5}$ for the ISM scenario and $\log[(E_\mathrm{jet}/A_\star)\,\mathrm{erg^{-1}}]=51.2\pm0.2$ in the wind case, respectively (median values with $1\sigma$ credible intervals). The value for the wind scenario is in good agreement with those reported in the literature. For the ISM scenario, although our estimate is larger than those previously reported, likely due to the combined action of the different definition of FWHM and the fact that we did not include the spectral information in our fit, it is still compatible with the estimates in the literature within $3\sigma$. 

According to our results, the collimation-corrected Sedov ratio of GRB\,030329 is somewhat lower than that of GRB\,221009A, with a mismatch at around 2.2 sigma for the ISM case, and 1.8 sigma for the wind case. This reflects the fact that the rest-frame size evolution of GRB\,030329, while similar to that of GRB\,221009A, is systematically offset towards smaller sizes and slower expansion speeds (see Fig.\ \ref{fig:size_beta_vs_time}).

The ratio of the collimation-corrected Sedov lengths of the two bursts, which is somewhere between 2 and 86 (union of 1-sigma intervals on the ratios in the ISM and wind cases), can be contrasted with that of their isotropic-equivalent gamma-ray energies, which is about $10^3$ \citep{lesage2023,Kaneko2007}. This points either to a large difference in the external medium densities, with GRB\,221009A exploding in a much denser environment than GRB\,030329, or to a different degree of collimation, with GRB\,221009A being much more collimated than GRB\,030329. Afterglow modelling \citep[e.g.][]{lhaaso2023,OConnor2023,Gill2023} does suggest the presence of a very narrow core in the GRB\,221009A jet, but it also points to shallow `wings' that contain even more energy than the core, thus contradicting the latter explanation.

\begin{figure*}
\centering    \includegraphics[width=0.7\textwidth]{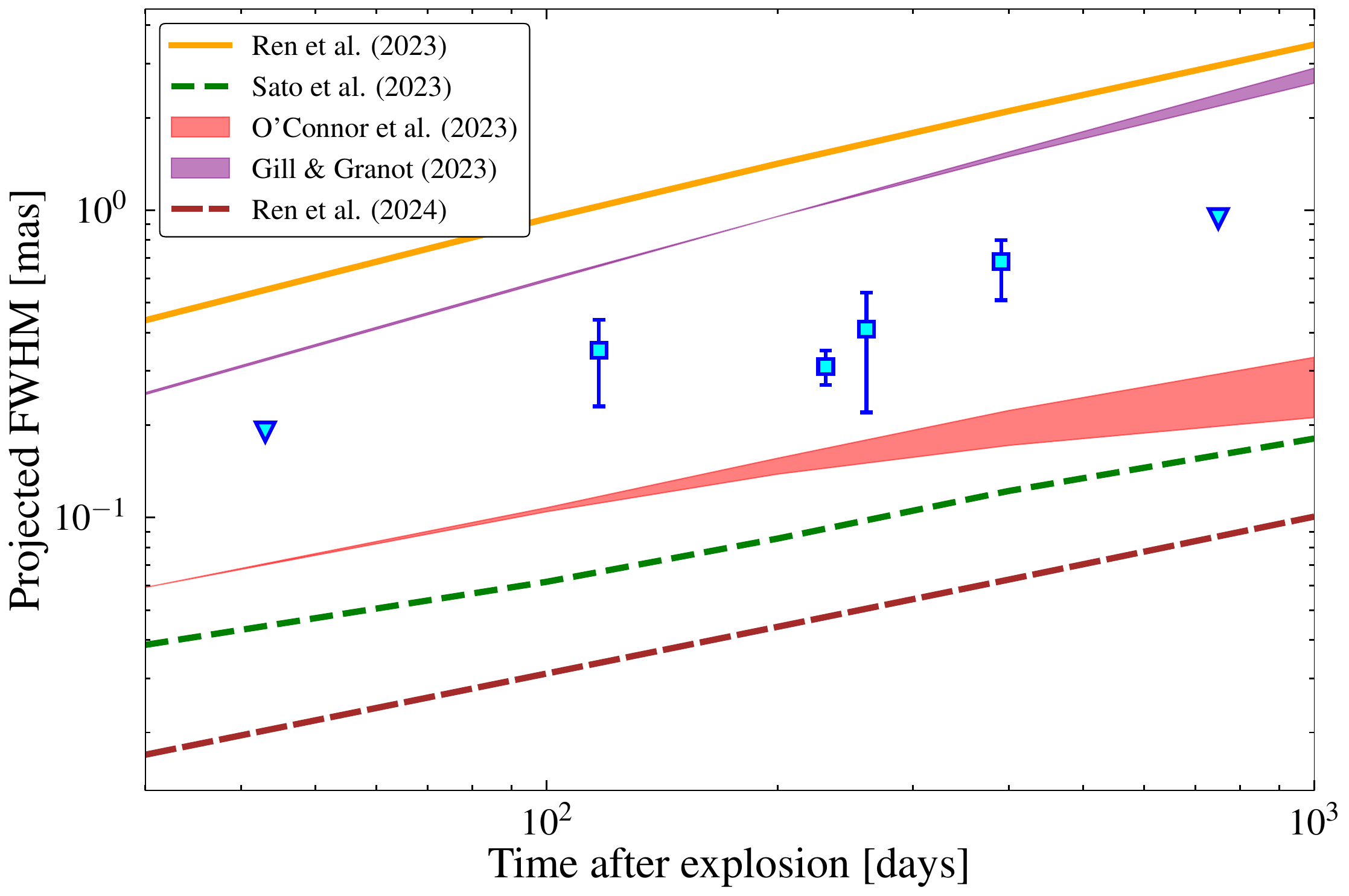}
    \caption{Comparison between the observed size evolution of GRB\,221009A and the prediction of the afterglow models proposed in the literature. The blue squares show the VLBI observations at 5\,GHz, while the predicted angular sizes of the models derived with \texttt{VegasAfterglow} are displayed with colours and styles that follow the legend of Fig.\,\ref{fig:Ejet_hist}.}
    \label{fig:models_vs_obs}
\end{figure*}

\subsection{Size evolution}

While informative, our collimation-corrected Sedov ratio comparison is not fully self-consistent, because of the differences in the assumed jet structures. Here we extend the comparison by using \texttt{VegasAfterglow} to predict the size evolution entailed by the models in the literature and perform a direct comparison with our observations. We adopted the physical parameters and jet structures reported in the corresponding studies and verified that the light curves and spectra produced by \texttt{VegasAfterglow} reproduce the published results. The images were generated using the \texttt{sky\_image} attribute of the \texttt{model} class with a field of view of 20\,milliarcsecond and 8192 pixels, to ensure adequate angular resolution and map extent. Each image was visually inspected to confirm that the surface brightness distribution was properly sampled and to compare the computed FWHM with the image. The size evolution of the models was derived following the same procedure as that described in Sec.\,\ref{sec:physical_model}. Although we checked the consistency of the published light curves and spectra with the results of \texttt{VegasAfterglow}, we note that we could not verify whether the surface brightness distribution predicted by each model at a given time agrees with the one generated with \texttt{VegasAfterglow}, given that we do not have access to the original codes of the published models. 

\cite{Ren2023} modelled the early multi-wavelength observations, from the very high energies to the radio band, assuming a top-hat jet expanding into a cold external medium with a wind-like density profile. The model includes the contribution from a single FS, accounting for both the synchrotron and the SSC emission. 
As shown in Figure \ref{fig:models_vs_obs}, the size evolution predicted by the model (solid orange line) systematically overshoots the observed angular size.

\cite{Sato2023} proposed a two-component jet expanding into an ISM. Their model consists of a narrow, highly energetic and ultra-relativistic core surrounded by wider, less energetic, and slower `wings'. Their computation of the emission includes both synchrotron and SSC from the FS. 
We generated images for each jet component separately and combined their surface brightness distributions at each epoch before measuring the FWHM. The resulting angular-size evolution is shown in Fig.\,\ref{fig:models_vs_obs} (green dashed line). The model systematically underpredicts the observed angular size. The radio emission is dominated by the narrow core until approximately 100\,days after the burst. However, even when considering only the wider component, the angular sizes predicted by \texttt{VegasAfterglow} remain significantly smaller than the observations.

\cite{OConnor2023} proposed a structured jet with a shallow energy angular profile expanding in the ISM. The authors suggested that the afterglow emission could be reproduced by a combination of FS from this structured jet and RS, which they model as spherical for simplicity. Specifically, the early radio emission is ascribed to the RS, while the emission beyond approximately $200$\,days post-burst is dominated by the contribution from the FS. Since our global-VLBI observations probe this late-time regime, we considered only the FS component, which also facilitates the implementation of the model in \texttt{VegasAfterglow}. We also note that the physical parameters of the RS were not reported, preventing us from modelling this component. We adopted the angular energy profile given by their Eq.\,3. 
Figure\,\ref{fig:models_vs_obs} shows the predicted angular size of the model (red area, which brackets the predictions obtained when varying the jet half-opening angle within the allowed range): the model systematically underpredicts the observed VLBI angular sizes by approximately a factor of 2.

\cite{Gill2023} also proposed a shallow angular structure of the jet energy profile. The authors considered a jet with a narrow core, outside of which the angular structure of the energy and the bulk Lorentz factor follows a power-law profile, propagating into a wind-like environment. Their model includes synchrotron emission from both the FS and the RS. We defined the energy and velocity profile of the jet following their Eq.\,2. 
We note that the RS emission predicted by \texttt{VegasAfterglow} for the same parameters is markedly different from that shown in the figures of \citet{Gill2023}. Nevertheless, the FS emission is consistent with their results and, given that the radio emission beyond $200$\,days post-burst is dominated by the FS component, we focused our analysis on this contribution.
Figure\,\ref{fig:models_vs_obs} shows the predicted angular size of the model (purple area, which brackets the predictions obtained when varying the jet half-opening angle within the allowed range), which overshoots the observed size measurements by approximately a factor of 2.

\cite{Ren2024} proposed a structured jet characterised by a core surrounded by wings, expanding into an external medium with a density that transitions from an ISM-like profile in the region close to the progenitor, to a wind-like profile at larger distances from the burst, with the transition occurring at $r_\mathrm{tr}\sim 4\times 10^{16}\,\mathrm{cm}$. The model includes synchrotron radiation from both the FS and the RS, as well as SSC emission from the jet core. We defined the energy, Lorentz factor and medium density profiles in \texttt{VegasAfterglow} following their Eqs.\,1--3. Since the code does not allow different values of the microphysical parameters for different regions of the same jet, we adopted the FS parameters of the wings, which provide the dominant contribution at 5\,GHz at the times sampled by our observations, as can be inferred from the figures in \citet{Ren2024}. 
Fig.\,\ref{fig:models_vs_obs} shows the predicted angular size of the model (brown long-dashed line), is systematically below the observed size measurements. We note that \cite{Ren2024} suggested that the radio emission could be dominated by a slower ejecta cocoon. Although the inclusion of this component is expected to result in a larger predicted angular size of the source, it is not included in their model and therefore we could not test it.

\subsection{Other proposed models}
As shown in the previous section, the models proposed in the literature that can be verified with \texttt{VegasAfterglow} struggle at reproducing the observed angular size evolution at 5\,GHz.

Other models were proposed to explain the broad-band emission of GRB\,221009A. However, for most of them we could not reproduce their light curves and spectra with \texttt{VegasAfterglow}, and hence did not proceed with a direct comparison. Therefore, we are not able to test these models. In this section we briefly describe them and their features. 

\cite{Laskar2023} considered a top-hat jet propagating into a wind-like environment to explain the afterglow observations up to 76\,days post-burst. Their model, which includes the synchrotron emission of a single FS, is unable to reproduce the radio and millimetre afterglow emission. Although a single FS does not account for the radio data before $\sim100$\,days post-burst, this component is expected to dominate at late times, when our VLBI observations were obtained. To test this possibility, we compared the expected light curves and spectra obtained with \texttt{VegasAfterglow} assuming the parameters of the proposed model with our observations. The predicted flux densities are inconsistent with our late time observations. Therefore, we could not include this study among our direct comparisons.

\cite{Zheng2024} investigated a two-component jet model, including a narrow uniform core and wide structured wings. The flux densities predicted by \texttt{VegasAfterglow} with their input parameters are around an order of magnitude fainter, in the radio band (especially at 1.3\,GHz), than those shown in the figures of the article. Since we were unable to reproduce the proposed model, we could not include it among our direct comparisons.


Fitting the X-ray and the radio observations with a phenomenological fit, \cite{Rhodes2024} proposed a scenario where the synchrotron emission comes from three regions: external material shocked by a FS, GRB ejecta shocked by a RS and a shocked cocoon. Since the authors do not derive the physical parameters of the shocks, we could not reproduce their model with \texttt{VegasAfterglow}. 

\citet{Foffano2024} and \citet{Fraija2024} proposed a temporal evolution of the micro-physical parameters to explain the afterglow of GRB\,221009A. However, none of the proposed models considered the radio band. Therefore, we could not include these models among our direct comparisons.

\cite{Yao2025} favoured a single FS from a top-hat jet expanding into the ISM.  
In contrast, \cite{Zhang2025} proposed a two-component jet consisting of a narrow top-hat core dominated by the magnetic energy and a wide, matter-dominated structured jet. The model considers the electron synchrotron and SSC emission from the FS of the two-component jet, and the proton synchrotron emission from the RS of the wide jet, to explain the very high energy emission. A two-component jet, propagating into the ISM, was also employed by \cite{Sato2025}, who accounted for synchrotron and SSC radiation from a FS. Since we were not able to reproduce the light curves and spectra of these models, with \texttt{VegasAfterglow} systematically underpredicting the flux density in the radio band, we could not include them in our analysis.

Finally, to explain the afterglow light curves, spectra and the size measurements with VLBI observations at 15\,GHz, \cite{Geng2025} proposed a complex scenario in which a late, faster shell catches up with an early, slow shell. The interaction of the two shells generate a FS and a RS. The model also considers the transition from a wind-like environment to the ISM. Since \texttt{VegasAfterglow} cannot simulate such a proposed scenario, we could not include this model in our analysis.

\section{Conclusions}
\label{sec:conclusions}

In this work, we presented our global-VLBI observations of GRB\,221009A at 5\,GHz. The high quality of the dataset, spanning three epochs from 231 to 750\,days post-burst, enabled the measurement of the projected angular size of the GRB afterglow with a precision of 13\% and 30\% at 231 and 391\,days post-burst. We confirm the size expansion previously reported in the literature.
We fitted our observations, together with publicly available size measurements at 5\,GHz, with a single FS component using the public code \texttt{VegasAfterglow}, within a Bayesian framework. We considered both a homogeneous external medium and a surrounding medium with a wind-like density profile. The fit yielded a relatively well-constrained value of the logarithm of the collimation-corrected Sedov ratio, namely
$\log[(E_\mathrm{jet}/n)\,\mathrm{erg}^{-1}\,\mathrm{cm}^{-3}] = 54.2^{+0.4}_{-0.5}$ for the ISM scenario and $\log[(E_\mathrm{jet}/A_\star)\,\mathrm{erg}^{-1}] = 51.7^{+0.4}_{-0.2}$ for the wind-like scenario (median values with $1\sigma$ credible intervals). 

We employed \texttt{VegasAfterglow} to investigate whether the afterglow models proposed in the literature to explain the light curves and spectra of GRB\,221009A can also account for the observed evolution of the angular size at 5\,GHz. Among the proposed models that we were able to reproduce with \texttt{VegasAfterglow}, none can successfully reproduce the VLBI measurements. 
Therefore, our work confirms that the standard afterglow model currently struggles to explain the radio emission, not only in terms of flux density measurements (see, e.g., \citealt{Laskar2023, Rhodes2024}), but also in terms of source size, further corroborating the importance of VLBI observations of GRB afterglows as complementary probe for their dynamics, structure, and emission mechanisms.

Notwithstanding the high S/N of our dataset, only the first two global-VLBI observations achieved the $(uv)$--plane coverage required to obtain a relative uncertainty on the FWHM of $\lesssim 30\%$. The lack of additional high S/N measurements limited our ability to better constrain the temporal evolution of the projected angular size of GRB\,221009A. Although the expansion is confirmed at a $4.5\sigma$ confidence level, the large uncertainties in the temporal slope of the expansion prevent us from distinguishing between the ISM and wind-like scenarios, as well as between pre- and post-jet-break evolution. Nevertheless, our data suggest a preference for a post-jet-break expansion and, in the ISM case, for a significant lateral expansion. Overall, our results point out the importance of robust size measurements, which currently require the sensitivity and angular resolution provided by the global-VLBI network. The advent of the next generation VLA (ngVLA) and the inclusion of the future Square Kilometer Array (SKA) in a VLBI array will represent a major advance. Specifically, they will significantly improve both the sensitivity and the angular resolution, particularly in declination, allowing for precise and robust size measurements \citep{Murphy2018, Wilner2024, Giarratana2026}.

\begin{acknowledgements}
The European VLBI Network is a joint facility of independent European, African, Asian, and North American radio astronomy institutes. Scientific results from data presented in this publication are derived from the following EVN project codes: GG087, GG088.
The National Radio Astronomy Observatory is a facility of the National Science Foundation operated under cooperative agreement by Associated Universities, Inc. 
The Long Baseline Array is part of the Australia Telescope National Facility (\url{https://ror.org/05qajvd42}) which is funded by the Australian Government for operation as a National Facility managed by CSIRO.

SG thanks I. Mariani for her support in the visualization and graphical preparation of the radio maps.

SG acknowledges financial support from the Istituto Nazionale di Astrofisica (INAF) for the RAdio to TeV Transient Sources (RATTS) funding, project code: C53C22002020001. SG acknowledges financial support from the Istituto Nazionale di Astrofisica (INAF) for the ``Bando di Astrofisica Fondamentale 2024'', project number: 1.05.24.07.04. This work has been funded by the European Union-Next Generation EU, PRIN 2022 RFF M4C21.1 (202298J7KT - PEACE). This research acknowledges support from the Italian Space Agency (Agenzia Spaziale Italiana, ASI),  through contract ASI n. 2025-5-U.0: ``Gamma-ray bursts: a probe of multi-messenger and extreme Astrophysics''. This work has been funded by the ``GRB PrOmpt Emission Modular Simulator'' (POEMS) project, financed by INAF Grant 1.05.23.06.04. BM acknowledges financial support from the State Agency for Research of the Spanish Ministry of Science and Innovation, and FEDER, UE, under grant PID2022-136828NB-C41/MICIU/AEI/10.13039/501100011033; an Advanced Grant from the European Research Council (ERC) under the European Union’s Horizon 2020 research and innovation programme (`EuroFlash’; Grant agreement No. 101098079);  an NWO-Vici grant (`AstroFlash'; VI.C.192.045); and a Canada Excellence Research Chair in Transient Astrophysics (CERC-2022-00009).
JKL acknowledges support from the University of Toronto and Hebrew University of Jerusalem through the University of Toronto - Hebrew University of Jerusalem Research and Training Alliance program. MPT acknowledges financial support from the Severo Ochoa grant CEX2021-001131-S and from the Spanish grant PID2023-147883NB-C21, funded by MCIU/AEI/ 10.13039/501100011033, as well as support through ERDF/EU.

\end{acknowledgements}

\bibliographystyle{aa}
\bibliography{bibliography}

\begin{appendix}

\section{Consistency check of the FWHM measurements}
\label{appendix:check_FWHM}
An inconsistency between the definitions of the model FWHM and the actual measured FWHM would result in a substantial difference in the size computed from the models in \texttt{VegasAfterglow} on the one hand, and the real size predicted by the models on the other hand. This would, in turn, nullify our attempt to test the models proposed in the literature with the GRB size as observed with the VLBI. To check the consistency between the two definitions, we performed a series of simulations as follows. 
First, we generated the surface brightness distribution of different models, with different physical parameters, angular structures and distances. The redshift was chosen in order to have the GRB afterglow in the `super-resolution' regime, namely the source should not appear resolved in the de-convolved radio image. We changed pixel size and field of view of the synthetic map in order to ensure that the image was well-resolved, and to assess the potential impact of their choice. For each realisation, we produced the surface brightness distribution with \texttt{VegasAfterglow}, and computed the model FWHM using Eq.\,\ref{eq:fwhm}.

We then simulated an observation of the GRB afterglow with the global-VLBI array with the \texttt{eht-imaging} software. We produced synthetic visibilities of the observation with the \texttt{observe} task in \texttt{eht-imaging}, considering the standard frequency setup employed in our global-VLBI observations, with central frequency of 5\,GHz and a bandwidth of 512\,MHz. We assumed a standard phase-referencing experiment of 24\,hr, with a cycle time of 6 minutes: 4 minutes on the target, and 2 minutes on the phase calibrator. The GRB afterglow was represented by injecting the surface brightness distribution previously generated with \texttt{VegasAfterglow}, through the \texttt{load\_txt} attribute. Baseline-dependent thermal noise was added with the \texttt{add\_th\_noise} parameter. Information on each antenna, namely the geocentric coordinates of the station and the system equivalent flux densities (SEFDs, in Jy) were taken from the EVN Observation Planner\footnote{\url{https://github.com/bmarcote/vlbi_calculator/blob/7a3b8c5cbb8fc0ab2d1598a5b7a85200794e5df4/src/vlbiplanobs/data/stations_catalog.inp}}. The position of GRB\,221009A was assumed as phase centre. We imposed that each antenna be able to observe the source only when it is at an elevation between 20$^{\circ}$ and 80$^{\circ}$, as seen from the telescope site.
The synthetic visibilities were then imported in \texttt{Difmap}, and the projected angular size was derived with the \texttt{modelfit} task. To verify the impact of different realisations of the r.m.s. noise level generated by \texttt{eht-imaging}, we also fixed the physical and geometrical parameters, and simulated the same image different times. Finally, we compared the size estimates obtained with Eq.\,\ref{eq:fwhm} and in \texttt{Difmap}. We found that, in the `super-resolution' regime sampled by our VLBI observations, the two estimates differ by less than approximately 10\% of the size. When the projected angular size becomes comparable to the beam-size of the observation, the two estimates differ up to 20\% or more. However, the observed projected angular size of our VLBI observations is always below the diffraction limit of the array. Moreover, we found that the pixel size and the random noise of the image contribute for 2\% and 5\%, respectively, to the difference between the two estimates.

\end{appendix}
\end{document}